\documentclass[12pt,a4paper]{article}            
 \usepackage[skins,theorems]{tcolorbox}
\tcbset{highlight math style={enhanced,
  colframe=red,colback=white,arc=0pt,boxrule=1pt}}
  \usepackage[bookmarksopen, bookmarksnumbered, bookmarksopenlevel=2]{hyperref}
  \usepackage{tikz}
  \usepackage{tikz-3dplot}
 \usetikzlibrary{calc}
 \usetikzlibrary{decorations} %
 \usepackage[UKenglish]{babel}
 \usepackage[toc,page]{appendix}
 \usepackage{amsmath}
 \usepackage{amssymb}
 \usepackage{graphicx}
 \usepackage{hhline}
 \usepackage[bf]{caption}
\usepackage{cite}
\usepackage[vcentermath]{youngtab}
\usepackage{geometry}
\usepackage{slashed}
\usepackage{color}
\usepackage{stackrel}
\usepackage{tikz-cd} 
\usepackage{mathtools}
\usepackage{cancel} 
\usepackage{multirow}
\usepackage[margin=0pt,font=small,labelfont=normalfont,skip=22pt]{subcaption}

\usepackage{empheq}
\usepackage{arydshln}

\newcommand{\bqa}{\begin{eqnarray}}
\newcommand{\eqa}{\end{eqnarray}}

\newenvironment{eqn*}{\begin{equation*}\begin{aligned}}{\end{aligned}\end{equation*}\noindent}
\hypersetup{
    pdftitle={},
    pdfauthor={},
    pdfsubject={}
}
\numberwithin{equation}{section}
\numberwithin{table}{section}

\makeatletter

\newcommand{\be}{\begin{equation}}
\newcommand{\ee}{\end{equation}}
\newcommand{\beq}{\begin{equation}}
\newcommand{\eeq}{\end{equation}}
\newcommand{\ba}{\begin{aligned}}
\newcommand{\ea}{\end{aligned}}

\newcommand{\bea}{\begin{eqnarray}}
\newcommand{\eea}{\end{eqnarray}}

\newcommand{\cN}{\mathcal{N}}

\newcommand{\cF}{\mathcal{F}}

\newcommand{\cV}{\mathcal{V}}

\newcommand\bi{\begin{itemize}}
\newcommand\ei{\end{itemize}}

\def\unit{{1\kern-.65ex {\rm l}}}
\def\1{{1\kern-.65ex {\rm l}}}

\def\bbF{{\mathbb{F}}}

\newcount\hour \newcount\minute
\hour=\time \divide \hour by 60
\minute=\time
\def\now{%
\ifnum \hour<13
  \ifnum \hour=0 \advance \hour by 12 \number\hour:\else \number\hour:\fi%
     \ifnum \minute<10 0\fi%
     \number\minute%
\ A.M.%
\else \advance \hour by -12 \number\hour:%
  \ifnum \minute<10 0\fi%
  \number\minute%
  \ P.M.%
\fi%
}

\makeatother

\begin{document}

\begin{titlepage}
\begin{center}
\rightline{\small }

\vskip 15 mm

{\large \bf
 Moduli-dependent Species Scale 
} 
\vskip 11 mm

Damian van de Heisteeg,$^{1}$ Cumrun Vafa,$^{2}$ Max Wiesner,$^{1,2}$ David H. Wu$^{2}$

\vskip 11 mm
\small ${}^{1}$ 
{\it Center of Mathematical Sciences and Applications, Harvard University,\\ Cambridge, MA 02138, USA}  \\[3 mm]
\small ${}^{2}$ 
{\it Jefferson Physical Laboratory, Harvard University, Cambridge, MA 02138, USA}

\end{center}
\vskip 17mm

\begin{abstract}
The counting of the number of light modes in a gravitational theory is captured by the notion of the `species scale', which serves as an effective UV cutoff below the Planck scale.  We propose to define a moduli-dependent species scale in the context of 4d,  ${\cal N}=2$ theories, using the one loop topological free energy $F_1$, which we relate to a gravitational version of the $a$-function.  This leads to $\Lambda_{\rm sp}\sim \frac{1}{\sqrt{F_1}}$ from which we recover the expected scaling of the species scale in various corners of the moduli space.  Moreover by minimizing $F_1$ we define the center of the moduli space (the `desert point') as a point where the species scale is maximal.  At this point the number of light degrees of freedom is minimized.
\end{abstract}

\vfill
\end{titlepage}

\newpage

\tableofcontents

\setcounter{page}{1}
\section{Introduction}
The Planck scale $M_{\rm pl}$ serves as a fundamental UV cutoff in gravitational theories.  However, when there are many light degrees of freedom, a lower scale called the `species scale' $\Lambda_{\rm sp}<M_{\rm pl}$ plays that role \cite{Dvali:2007wp,Dvali:2007hz,Dvali:2010vm,Arkani-Hamed:2005zuc, Distler:2005hi, Dimopoulos:2005ac, Dvali:2009ks,Dvali:2012uq}. It is defined by the condition that $R_{\rm min}\sim 1/\Lambda_{\rm sp}$ is the radius of the smallest black hole for which a quarter of the area of the horizon gives the number of degrees of freedom.  Clearly, when there are many light modes, the black hole entropy would be larger as they can be used as parts of the degrees of freedom of the black hole, and this leads to the identification $N_{\rm sp}\sim \left( M_{\rm pl} R_{\rm min}\right)^{d-2}\sim \left(\frac{M_{\rm pl}}{\Lambda_{\rm sp}}\right)^{d-2}$.

In the context of supersymmetric theories, we often end up with exactly massless fields, called the moduli fields. In these cases, one would expect that $\Lambda_{\rm sp}$ could depend on the value of these moduli fields. Indeed, it is known that as we change the moduli field vevs the masses of the states change and thus the number of light degrees of freedom, $N_{\rm sp}$, would depend on them and thus so does $\Lambda_{\rm sp}$. This raises the question of whether we can give a computable formulation of the moduli dependence of $\Lambda_{\rm sp}$.  The aim of this note is to provide such a notion for the vector multiplet moduli for ${\cal N}=2$ supergravity theories in four dimensions.

This general question was studied recently in \cite{Long:2021jlv} where the idea pursued was to study the gap in the spectrum of massive states to define an analog of the species scale, i.e.~where the first massive state, not part of the EFT, appears.  Moreover since finding the full spectrum of the gravitational theory as a function of moduli is too difficult, the attention was directed to the BPS excitations to define this notion.
In the present note, we consider the special case of 4d, ${\cal N}=2$ supergravity theories and improve this in two ways: First we do not restrict our attention only to the BPS states; and secondly, the notion of the species scale we formulate is more naturally connected with the definition of the species scales as counting the number of light degrees of freedom.  We argue that the one loop topological string free energy $F_1$ is a measure of light species, which we motivate by connecting it to a gravitational version of the $a$-function.  Furthermore, we check that at the corners of moduli space (such as large distance points and near conifold points) it leads to the expected behavior.  Moreover, we use this to compute the point with the minimal number of light species by minimizing $F_1$.  This can be viewed as a `desert point' where not many light degrees of freedom exist beyond the EFT until we reach the UV cutoff.  For example, for the mirror of the quintic threefold, this leads to the Landau-Ginzburg point as the desert point.
We also ask whether this notion of a desert point agrees with the BPS notion defined in \cite{Long:2021jlv}.  We find that in some cases (such as for the quintic) it does and in some other cases it does not.

The organization of this paper is as follows: In section~\ref{sec:BPSreview} we review the notion of species scale and its dependence on the moduli including defining `the center of the moduli space' in terms of the species scale. In section \ref{sec:F1review} we review some properties of the genus one free energy, $F_1$, for topological strings. In section~\ref{sec:3} we argue that $F_1$ can be used to measure the number of species and test our proposal at asymptotic points in moduli space. In section~\ref{sec:ex} we discuss explicit examples and identify the points in moduli space where $F_1$ is minimized and compare to the points where the BPS mass gap is maximized. We present our conclusion in section~\ref{sec:conclusions}. The appendix contains some additional details on the examples discussed in section~\ref{sec:ex}.

\section{Species scale and the desert}
\label{sec:BPSreview}
Let us consider a low-energy effective theory in $d$-dimensions coupled to gravity. For such a theory, we can define two scales: First the EFT cut-off  scale, $\Lambda_{\text{\tiny EFT}}$, which corresponds to the scale of the first massive excitation not described by the EFT. On the other hand, there is the quantum gravity scale, $\Lambda_{\text{\tiny QG}}$, at which gravity becomes strongly coupled and the low-energy EFT description break down. As mentioned in the introduction, this latter scale can be identified with the species scale \cite{Dvali:2007wp,Dvali:2007hz,Dvali:2010vm,Arkani-Hamed:2005zuc}
\begin{equation}
    \Lambda_{\rm sp} = \frac{M_{\rm pl}}{N^{\frac{1}{d-2}}}\,, 
\end{equation}
which equivalently can be viewed as the mass scale whose inverse sets the radius of the smallest black hole that can be reliably described in the low-energy EFT. In order to compute $\Lambda_{\text{\tiny EFT}}$ or $\Lambda_{\rm sp}$ one needs the data of massive states in the theory. In a theory of gravity all parameters are expected to be given by the vevs of scalar fields, $\phi$, such that the mass of states depends on the value for their vevs $\langle \phi\rangle$. In the following we consider the case that the scalar fields are exactly massless such that their vevs describe the moduli space of the theory. As the masses of the massive states of the theory depend on the vevs of the scalar fields, so do the scales $\Lambda_{\text{\tiny EFT}}$ and $\Lambda_{\text{\tiny QG}}$. The Distance Conjecture \cite{Ooguri:2006in} predicts that as we approach an infinite-distance boundary moduli space there exists a tower of states that becomes massless in Planck units exponentially fast in the moduli space distance. Therefore in such a limit $\Lambda_{\text{\tiny EFT}}, \Lambda_{\rm sp}\rightarrow 0$. On the other hand in the interior of moduli space both $\Lambda_{\text{\tiny EFT}}$ and $\Lambda_{\rm sp}$ are finite. In particular, there need to be points in moduli space where either $\Lambda_{\text{\tiny EFT}}$ or $\Lambda_{\rm sp}$ are maximized. As these points can be thought of having the least amount of light states they should be viewed as far away from the boundaries of moduli space (where both cut-off scales vanish) and therefore they are natural candidates for a definition of the center of the moduli space. In this context, there are hence two possible definitions of the center of the moduli space as the point where $a)$ $\Lambda_{\text{\tiny EFT}}$ is maximized or $b)$ $\Lambda_{\rm sp}$ is maximized. 

In order to find the center of the moduli space according to definition $a)$ one would need to know the entire mass spectrum of the theory. Even in theories with a large number of supercharges, e.g. $\cN=2$ in $d=4$, this is a rather difficult task since only the masses of BPS states can be computed reliably. Still, as a starting point instead of $\Lambda_{\text{\tiny EFT}}$ one may consider the BPS mass gap, $\Lambda_{\text{\tiny BPS}}$, corresponding to the mass of the lightest BPS state in the spectrum. The behavior of $\Lambda_{\text{\tiny BPS}}$ in the interior of moduli space was investigated in \cite{Long:2021jlv}. 

In the presence of non-BPS states $\Lambda_{\text{\tiny BPS}}$ can differ significantly from $\Lambda_{\text{\tiny EFT}}$ (and $\Lambda_{\rm sp}$). To illustrate this, consider e.g.~the heterotic $E_8\times E_8$ string in ten dimensions and its eleven-dimensional lift given by M-theory on the interval $S^1/\mathbb{Z}_2$ \cite{Horava:1995hw}. In this setting the only BPS objects are the fundamental string and the NS5-branes whereas non-BPS states arise e.g.~from the KK modes on the M-theory interval. The single modulus corresponds to the heterotic string coupling $g_H$ which is related to the length $r_H$ of the M-theory interval as 
\begin{equation}
    g_H = (r_H/l_{11})^{3/2}\,,
\end{equation}
where $l_{11}$ is the 11-dimensional Planck length. In terms of $g_H$ the mass scales set by the BPS objects are given by 
\begin{equation}
    \frac{M_{\text{\tiny F1}}}{M_{\rm pl,10}} = g_H^{1/4}\,,\qquad  \frac{M_{\text{\tiny NS5}}}{M_{\rm pl,10}}= \frac{1}{g_H^{1/12}}\,. 
\end{equation}
For instance, in the strong coupling limit $g_H\rightarrow \infty$ the BPS mass gap is thus given by 
\begin{equation}
    \frac{\Lambda_{\text{\tiny BPS}}}{M_{\rm pl,10}} = \frac{1}{g_H^{1/12}}\,.
\end{equation}
However, the actual mass gap in this limit is set by the non-BPS KK modes scaling like 
\begin{equation}
    \frac{\Lambda_{\text{\tiny EFT}}}{M_{\rm pl,10}} = \frac{M_{\text{\tiny KK}}}{M_{\rm pl,10}} = \frac{1}{g_{H}^{3/4}} \ll \frac{\Lambda_{\text{\tiny BPS}}}{M_{\rm pl,10}} \,,
\end{equation}
in the limit $g_H\rightarrow \infty$. We thus see that $\Lambda_{\text{\tiny BPS}}$ is not necessarily a good measure for the mass-gap in a given theory. Since the exact mass spectrum of a given theory is difficult to determine and so $\Lambda_{\text{\tiny EFT}}$ is hard to compute, in this note we focus on the definition of the center of the moduli space as the point where $\Lambda_{\rm sp}$ is maximized. 

Still, in order to make practical use of this definition, we need a way to encode the moduli dependence of the species scale. Whereas in asymptotic limits in moduli space this can be done since the light spectrum and their moduli-dependent masses are well-known, to maximize $\Lambda_{\rm sp}$ we need to have a closed expression for the species scale valid all over moduli space.

In what follows, we will focus on compactifications of Type II string theories on Calabi--Yau threefolds, $Y_3$, which results in a 4d $\mathcal{N}=2$ supergravity theory. In particular, we propose that $\Lambda_{\rm sp}$ is related to the genus-one free energy, $F_1$, of topological strings propagating on $Y_3$. Specifically, we will argue that 
\begin{equation}
    \frac{\Lambda_{\rm sp}}{M_{\rm pl}}\sim \frac{1}{F_1^{1/2}}\, ,
\end{equation}
holds in various regimes of the moduli space of $Y_3$ in Sec.~\ref{sec:3}. We argue for this by providing evidence that $N_{\rm sp}\sim F_1 $ from various viewpoints.
Furthermore, the maximum desert point in the moduli space is therefore captured by the minimum of the genus-one free energy of the topological string $F_1$.  

\section{Genus one free energy for topological strings} \label{sec:F1review}

Before arguing in the next section that the genus-one free energy can be used as a measure for the species scale, we first want to review some basic facts about the genus one free energy $F_1$ for topological strings due to its prominent role in identifying the center of the moduli space. Therefore, consider Type II string theories compactified on a Calabi--Yau threefold $Y_3$. The string worldsheet theory in this case can be described by an $\mathcal{N}=(2,2)$ CFT in two dimensions. For such a CFT, $F_1$ is defined as a moduli-dependent index \cite{Cecotti:1992vy, Bershadsky:1993cx}
\begin{equation}\label{F1index}
 F_1 = \frac{1}{2} \int_{\mathcal{F}} \frac{d^2 \tau}{\tau_2}  \text{Tr}\left[(-1)^F F_L F_R q^{H_0} \bar{q}^{\bar H_0} \right]\,,
\end{equation}
where $\mathcal{F}$ is the $SL(2)$ fundamental domain, $F_{L(R)}$ the left-(right-)moving fermion number, $H_0$ the 2d Hamiltonian and $q=e^{2\pi i \tau}$. The above integral diverges for the supersymmetric Ramond ground states of the theory. In order to get a convergent integral we should therefore subtract the ground state contribution to $F_1$. As a consequence, $F_1$ is only determined up to an additive constant. For say the topological B-model, one would naively expect $F_1$ to be the real part of a holomorphic function of the flat coordinates $t^i$ on the complex structure moduli space of the Calabi--Yau threefold since any $\bar{t}^i$ variation would be associated to an observable in the $\bar{\rm B}$-model, i.e. the model with complex conjugate twist. In the B-model the corresponding $\bar{\rm B}$-operators are BRST trivial such that one would not expect a $\bar{t}^i$ dependence of $F_1$. However, this is not quite true due to the holomorphic anomaly which imposes a mixing between the holomorphic and the anti-holomorphic sector. For $F_1$ the holomorphic anomaly equation reads \cite{Cecotti:1992vy}
\begin{equation}
\partial_{\bar \jmath}\partial_{i} F_1= \text{Tr}(-1)^F C_i \bar{C}_{\bar \jmath} -\frac{1}{12} G_{i\bar\jmath} \text{Tr}(-1)^F\,, 
\end{equation}
where $C_i$($\bar{C}_{\bar \jmath}$) are the structure constants of the (anti-)chiral ring of supersymmetric ground states and $G_{i\bar \jmath}$ is the Zamolodchikov metric on the field space. As shown in \cite{Bershadsky:1993ta}, the holomorphic anomaly equation can be integrated giving
\begin{equation}\label{F1general}
 F_1 = \frac{1}{2}\left(3+h^{1,1}-\frac{\chi}{12}\right) K\,-\frac{1}{2}\log \det G_{i\bar \jmath} +\log|f|^2\,,
\end{equation}
where $\chi$ is the Euler characteristic of $Y_3$, $K$ the K\"ahler potential, $G_{i\bar{\jmath}}$ the corresponding K\"ahler metric, and $f$ a holomorphic function that can be determined by studying the asymptotics of $F_1$ in the vicinity of the boundaries of the moduli space, cf.~\cite{Huang:2006hq} for a detailed discussion.

For instance, at large complex structure, i.e.~for $t^i,\bar{t}^i\rightarrow \infty$, the index in \eqref{F1index} can be determined directly leading to \cite{Bershadsky:1993ta}
\begin{equation}
F_1\stackrel{\hspace{-3pt}\text{\tiny LCS}}{\longrightarrow} \frac{1}{12} \int_{X_3} J\wedge c_2\,,
\end{equation}
where $X_3$ is the mirror of $Y_3$, $J$ its K\"ahler form and $c_2$ the second Chern class of $X_3$. In terms of the flat coordinates $t^i$ we thus have
\begin{equation}\label{asymptLCS}
F_1\stackrel{\hspace{-3pt}\text{\tiny LCS}}{\longrightarrow} \frac{1}{12} c_{2,i}\, \text{Re}\,t^i\,,
\end{equation}
with $c_{2,i}$ the expansion coefficients of $c_2$ in a basis of the K\"ahler cone.  On the other hand, in the vicinity of a conifold $F_1$ behaves as \cite{Vafa:1995ta} 
\begin{equation}\label{asymptconi}
 F_1 \stackrel{\hspace{-3pt}\text{\tiny coni}}{\longrightarrow} -\frac{1}{12} \log |\mu|^2+\dots \,,
\end{equation}
where the conifold is located at $\mu =0$ and the dots indicate sub-leading terms in the limit $\mu \rightarrow 0$. This can be understood by noticing that
at the conifold point a hypermultiplet becomes massless with mass $M/M_{\rm pl}\sim |\mu|$, leading to this divergent term in $F_1$ including the factor of $-1/12$ \cite{Vafa:1995ta}.\footnote{If at the conifold point instead of an $S^3$ a Lens space $S^3/G$ shrinks, the factor $1/12$ should be replaced by $|G|/12$ \cite{Gopakumar:1997dv}.} Given these asymptotics of $F_1$ close to the boundaries of the moduli space, one can completely fix the holomorphic function $f$. In general, $f$ has the form 
\begin{equation}\label{fansatz}
f(x) = x^{a_0} \prod_i \Delta_i^{a_i}\,.
\end{equation}  
Here $x$ is a coordinate on the moduli space such that $x=0$ corresponds to the LCS point where $x$ is related to the flat coordinate $t$ via  
\begin{equation}
 t = -\frac{1}{2\pi} \log x + \dots \,.
\end{equation} 
In addition, the $\Delta_i$ denote the components of the discriminant locus on the moduli space, and the exponents $a_0, a_i$ have to be fixed such that when plugging \eqref{fansatz} into \eqref{F1general} one recovers the asymptotic behavior of $F_1$ just discussed. 

Let us consider the form of the holomorphic ambiguity for the examples considered in this work. These complex structure moduli spaces have three singularities: a large complex structure point at $x=0$, a conifold point at $x=x_c$ and another singularity at $x=\infty$. The holomorphic ambiguity is fixed by requiring the correct asymptotic behavior at the first two singularities, so the type of the third singularity is irrelevant. To be precise, in a convenient coordinate system and K\"ahler gauge of the periods we find\footnote{As discussed in more detail in appendix \ref{app:periods}, we use the standard coordinate of the Picard-Fuchs equation, and for the resulting period vector we do not apply any rescalings by the first fundamental period.}
\begin{equation}
    f(x) = x^{-\frac{c_2}{24}-\frac{1}{2}}(x-x_c)^{-\frac{1}{12}}\, .
\end{equation}
The first factor assures that we have the right asymptotic behavior at the LCS point \eqref{asymptLCS} when $x\to 0$, where the minus half has been included to compensate for the asymptotic behavior of the K\"ahler metric; the second gives the correct asymptotic behavior at the conifold point \eqref{asymptconi} when $\mu = x - x_c \to 0$.

\section{Species scale and genus one free energy}
\label{sec:3}
In this section, we want to argue that the species scale $\Lambda_{\rm sp}$ of the $\cN=2$ gravity theory is related to the genus-one free energy. 
We will first give a heuristic argument based on the interpretation of $F_1$ in the B-model.  We then give another argument based on the analogy with the $a$-function in 4d CFTs.  Finally, we test our proposal at different boundaries in moduli space where the behavior of both $F_1$ and $\Lambda_{\rm sp}$ is known independently and see that it reproduces the expected results at these corners.

\subsection{\texorpdfstring{$F_1$}{F1 } and the spectrum of the Laplacian}

A formal measure of $N_{\rm sp}$ is given by the accumulation of the spectrum of the theory in the low mass region. Consider a compactification of Type IIB on a Calabi--Yau threefold. The spectrum of the Laplacian
$\Delta$ of the Calabi--Yau would give a measure of $N_{\rm sp}$. More specifically, if we consider $|{\rm log}(\det \Delta )|$, the growth of this quantity would be a measure for the accumulation of light mass modes.
This, however, has the deficiency that it is difficult to compute (not to mention the fact that it receives corrections by higher loop amplitudes).  Nevertheless, there is a combination of the Laplacian $\Delta_{(p,q)}$ acting on the $(p,q)$-forms
which turns out to be computable and is an index-like quantity. This is indeed given by the one loop topological string amplitude and is the combination \cite{Bershadsky:1993cx}
\begin{equation}\label{F1logdet}
    F_1= \frac{1}{2}\sum_{p,q}(-1)^{p+q}\left(p-\frac{n}{2}\right)\left(q-\frac{n}{2}\right)\ 
{\rm log}(\det \Delta_{(p,q)})\, ,
\end{equation}
where $n$ is the complex dimension of the target space of the topological string. Given that this is an index-like quantity it is possible that the light degrees of freedom cancel out between bosons and fermions. Indeed this happens if instead of a generic CY, we consider special ones, like $T^6$. However, for the more generic case we expect that $F_1$ will contain information about the light degrees of freedom of the theory.  
From this heuristic argument it is not clear with which  power of $N_{\rm sp}$ should $F_1$ scale.  In the remainder of this section we argue that $F_1$ should be proportional to $N_{\rm sp}$.
\subsection{Higher derivative corrections and the number of states}
To give a stronger and more quantitative argument for a relation between the genus-one free energy of topological strings and $\Lambda_{\rm sp}$ let us first recall that in the 4d effective action $F_1$ is related to the coefficients of higher derivatives correction
\begin{equation}\label{F14daction}
 S_{4d} \supset \int d^4 x \, \mathbb{F}_1  R_+^2\,,
\end{equation}
where we defined 
\begin{equation}\label{splitF1}
    \mathbb{F}_1= F_1(t, \bar t) + F_1^0\,,
\end{equation}
to account for the fact that the $F_1(t,\bar t)$ obtained from the topological string as in \eqref{F1general} provides the correct genus-one free energy up to an additive constant $F_1^0$ that is independent of the moduli $t, \bar{t}$. 
Here, $t$ are the flat coordinates on the vector multiplet moduli space and $R_+$ in \eqref{F14daction} is the self-dual part of the curvature tensor. In addition, we have a one-loop contribution to the 4d effective action yielding a four-derivative term\cite{Antoniadis:1993ze,Antoniadis:1997zt}: 
\begin{equation}\label{4dterm}
S_{4d} \supset \int d^4x\, \tilde{\mathbb{F}}_1 \,\frac{(\partial \partial S)^2}{2(\text{Im}\,S)^2}\,,
\end{equation}
where the field $S$ is related to the 4d dilaton, $\varphi_4$, as $\text{Im}\,S=\mathcal{V}_{Y_3}/g_s^2=e^{-2\varphi_4}$. Here $\tilde{\mathbb{F}}_1$ is another coefficient whose variation w.r.t. the scalars $z$ in the hypermultiplets can be determined from the topological string (at least at tree-level in $g_s$). Similar to \eqref{splitF1} we can therefore write 
\begin{equation}\label{splittildeF1}
  \tilde{\mathbb{F}}_1 = \tilde{F}_1(z,\bar z) +\tilde{F}_1^0\,,
\end{equation}
where now $\tilde{F}_1^0$ is independent of the scalars in the hypermultiplet.
It is interesting to note that the vector multiplet dependence of ${\mathbb{F}}_1$ is exact at the one loop level but the hypermultiplet dependence in \eqref{splittildeF1} receives non-perturbative corrections (consistent with the fact that the Type II string coupling belongs to a hypermultiplet). Having defined the coefficients of these higher-derivative terms, we may now ask how these are related to the number of light species.

Let us therefore digress briefly from the 4d effective supergravity obtained from the Type II string compactification and instead consider a spontaneously broken 4d CFT with a dilaton $\tau$ as Nambu-Goldstone boson. In the presence of curved background the theory typically develops a trace anomaly which is captured by anomaly-matching terms in the action (cf. \cite{Fradkin:1983tg,Komargodski:2011vj})
\begin{equation}\begin{aligned}\label{Sanomaly}
 S_{\rm anomaly} = &-a \int d^4 x \sqrt{-g} \left(\tau E_4 + 4(R^{\mu \nu} -\frac12 g^{\mu \nu} R)\partial_\mu\tau \partial_\nu \tau -4(\partial \tau)^2 \Box \tau +2 (\partial \tau)^4 \right) \\
&+c\int d^4 x\sqrt{-g}\tau W^2_{\mu \nu  \rho\sigma} \,.
\end{aligned}\end{equation} 
Here $E_4$ and $W_{\mu\nu\rho\sigma }$ are the Euler density and the Weyl tensor which can be expressed in terms of the Riemann tensor as 
\begin{equation}
E_4 =R_{\mu \nu \rho \sigma}^2 -4 R_{\mu \nu}^2 +R^2 \,,\qquad W_{\mu \nu \rho \sigma}^2 =R_{\mu \nu \rho \sigma}^2 -2 R_{\mu \nu}^2 +\frac13 R^2 \,.
\end{equation}
In addition, in \eqref{Sanomaly} we introduced the two coefficients $a$ and $c$. Of particular interest to us is the $a$-parameter which for a 4d CFT counts the number of degrees of freedom. The $a$-function is the coefficient of the four-derivative terms of $\tau$. On the other hand, in an $\cN=2$ supergravity theory we recall that $\tilde{\bbF}_1$ is the coefficient of the four-derivative term \eqref{4dterm} for the axio-dilaton. 

Now we come back to the case of supergravity theories.  In such a theory we do not have conformal symmetry (after all the Planck scale is always part of the gravity theory).
So there cannot be a precise relation to the trace anomaly of a conformal theory, which is scale invariant.  Nevertheless,
given that the dilaton $S$ is related in the same way to a rescaling of the CY-volume as $\tau$ is related to a rescaling of the spacetime volume, it is natural to identify $\exp(-2\tau) \leftrightarrow S$. Indeed, rescaling the 4d length scales by $\exp(-\tau)$ the Einstein term scales like $\exp(-2\tau)$. 
 On the other hand, for CY compactifications we get the Einstein term scaling as $\mathcal{V}_{Y_3}/g_s^2\sim {\rm Im}(S)$.  This motivates us to indeed identify 
\begin{equation}
  S \leftrightarrow \exp(-2\tau) \,,
\end{equation}
leading to
\begin{equation}
\Box S = - 2 S (\Box \tau)  + 4 S (\partial \tau)^2 \,,
\end{equation}
which allows us to rewrite the 4-derivative term in \eqref{4dterm} as 
\begin{equation}
 S_{\rm 4d}\supset 4 \int d^4 x \sqrt{g}\ \tilde{\mathbb{F}}_1(z,\bar z)  \left[4 (\partial \tau)^4 -4 (\Box \tau) (\partial \tau)^2 +  (\Box \tau)^2 \right]\,.
\end{equation}
which on shell (where $\Box \tau = (\partial \tau)^2$) is proportional to \eqref{Sanomaly}. This motivates us to  identify 
\begin{equation}\label{aF1identification}
 a\quad \longleftrightarrow \quad  \tilde{\mathbb{F}}_1\,. 
\end{equation}
Hence, $\tilde{\mathbb{F}}_1$ should count the `degrees of freedom' of the supergravity theory. More precisely the `a'-function counts the number of degrees of freedom of the massless fields in the CFT unlike the gravity case, which has a fixed Planck scale.  So in the same spirit we interpret $\tilde{\mathbb{F}}_1$ as counting the light degrees of freedom with mass much less than the Planck scale.  The ambiguity in exactly where we put this upper cutoff, may be related to the fact that $\tilde{\mathbb{F}}_1$ is only well defined up to an additive constant which can define a renormalization point for this upper scale.  A similar index-like object arises for the computation of stringy threshold corrections \cite{Kaplunovsky:1992vs} where it is shown that the ambiguity in the additive constant term there is related to the choice of the UV cutoff where the bare gauge coupling constant is defined.  With these motivations in mind,
we therefore propose $\tilde{\mathbb{F}}_1$ as a possible measure for the number of light degrees of freedom, i.e.
\begin{equation}\label{NspF1}
N_{\rm sp}\simeq \tilde{\mathbb{F}}_1\,.
\end{equation}  
Here '$\simeq$' indicates that $\tilde{\mathbb{F}}_1$ does in fact not give exactly the numbers of light degrees of freedom as it is an index for which there can be cancellation between bosons and fermions. In case there are systematic cancellations appearing in this index, $\tilde{\mathbb{F}}_1$ may thus be parametrically smaller than the actual $N_{\rm sp}$, as already discussed.  But generically we expect this relation to be valid. 

As mentioned above, for a given background we can only determine the dependence of $\tilde{\mathbb{F}}_1$ on the hypermultiplet sector up to a constant term. This constant can, however, depend on the chosen background and in particular on the vector multiplet moduli which for a given background are fixed to some value $t=t_0$. We may therefore update \eqref{splittildeF1} to allow for such a dependence 
\begin{equation}
    \tilde{\mathbb{F}}_1=\tilde{F}_1(z,\bar z) + \tilde{F}_1^0(t_0,\bar{t}_0)\,. 
\end{equation}
Notice that in the CFT case, the $a$-function is also the coefficient of the Euler density $E_4$ in the higher-derivative action which includes a Riemann-squared term. In the supergravity action the Riemann-squared term, on the other hand, has coefficient $\mathbb{F}_1$, cf. \eqref{F14daction}. The identification \eqref{aF1identification} then implies that the $t_0$-dependent term in $\tilde{\mathbb{F}}_1$ should be related to $F_1(t_0,\bar{t}_0)$ as 
\begin{equation}
    \tilde{F}_1^0(t_0,\bar{t}_0) = \alpha F_1(t_0,\bar{t}_0) + \text{const.}\,, 
\end{equation}
for some $\alpha>0$. Still, there is a constant piece reflecting the fact that e.g.~the zero modes are not captured by neither $F_1(t,\bar t)$ nor $\tilde{F}_1(z,\bar z)$. Allowing for variations of both vector and hypermultiplets we thus have 
\begin{equation}
    \tilde{\mathbb{F}}_1 = \tilde{F}_1(z,\bar z) + \alpha F_1(t,\bar t) + \text{const.}\,, 
\end{equation}
which, according to \eqref{NspF1}, counts the number of light states. Using
\begin{equation}
\frac{\Lambda_{\rm sp}}{\rm M_{\rm pl}} =\frac{1}{N_{\rm sp}^{1/2}}\,,
\end{equation}
we hence find 
\begin{equation}\label{proposal}
 \frac{\Lambda_{\rm sp}}{M_{\rm pl}} \sim \frac{1}{\tilde{\mathbb{F}}_1^{1/2}}\, .
\end{equation}
This is our proposal of a moduli-dependent species scale for ${\cal N}=2$ supergravity theories.  If we want to find the desert point on the moduli space, i.e.~to maximize $\Lambda_{\rm sp}$, we hence need to minimize both $F_1(t,\bar t)$ and $\tilde{F}_1(z,\bar z)$ individually. In the following, we mostly focus on the minimization of $F_1(t, \bar t)$, since unlike for $\tilde{F}_1(z,\bar z)$ there are no $g_s$ corrections to the result obtained from the holomorphic anomaly. Before we show this minimization in explicit examples, we first perform some consistency checks showing that \eqref{proposal} correctly reproduces the species scale in certain limits in vector multiplet moduli space where $\Lambda_{\rm sp}$ can be computed independently.

\subsection{Asymptotic checks}
To test whether \eqref{proposal} is sensible, we should check whether at asymptotic limits in moduli space, where $N_{\rm sp}\rightarrow \infty$, we reproduce this behavior using \eqref{proposal}.\footnote{In \cite{Stout:2022phm} it was argued for gravitational theories with a holographic dual that at infinite distance limits in the information metric \cite{Stout:2021ubb} the central charge of the dual CFT diverges. It would be interesting to understand the relation between this holographic count of degrees of freedom with our proposal here. } To that end, let us fix the hypermultiplet scalars at some value $z=z_0$ such that $\tilde{F}_1(z_0,\bar{z}_0)\ll \infty$ and tune the vector multiplet moduli to the vicinity of some singular point in moduli space at which $F_1(t,\bar t)$ diverges. Due to the divergence of $F_1(t,\bar{t})$ we have
\begin{equation}\label{proposalasymptotic}
    \tilde{\mathbb{F}}_1 \sim  F_1(t,\bar{t}) +\dots \,,\qquad \Rightarrow \qquad \Lambda_{\rm sp} \sim \frac{M_{\rm pl}}{\sqrt{F_1(t,\bar t)}}\,,
\end{equation}
where for simplicity we set $\alpha=1$. We reviewed the asymptotic behavior of $F_1$ at the LCS and conifold points in the section~\ref{sec:F1review}. Let us treat them separately to see how the scaling of $F_1$ reproduces the scaling of the species scale. 
\subsubsection{Large complex structure} 
As reviewed in the previous section, in the vicinity of the LCS point $F_1$ scales as 
\begin{equation}
 F_1 \sim \frac{1}{12} c_{2,i} t^i\,,
\end{equation} 
with $c_{2,i}$ the expansion coefficients of the second Chern class of the mirror ${X}_3$. For simplicity, let us concentrate on the single-modulus case. In this case $F_1$ is simply proportional to the single flat coordinate $t$. On the other hand, the limit $t\rightarrow \infty$ is dual to a decompactification limit from 4d to 5d \cite{Lee:2019oct}. To see this, consider the limit $t\rightarrow \infty$ in the dual vector multiplet moduli space of Type IIA compactified on ${X}_3$. In this setting we notice that in order to have a limit purely in the vector multiplet moduli space we need to keep the 4d dilaton $\mathcal{V}_{{X}_3}/g_s^2$ constant as it resides in a hypermultiplet. Since $\mathcal{V}_{ X_3}$ is cubic in $t$ keeping the 4d dilaton requires the scaling
\begin{equation}\label{gsscaling}
 g_s \sim t^{3/2}\,. 
\end{equation} 
The lightest state is now given by D0-branes, i.e. the KK modes of the decompactification from 4d to 5d with mass
\begin{equation}
    \frac{M_{\text{\tiny D0}}}{M_{\rm pl}} = \frac{1}{g_s} \left(\frac{g_s^2}{\mathcal{V}_{\tilde{X}_3}}\right)^{1/2} \sim \frac{1}{t^{3/2}}\,,
\end{equation}
where in the last step we used \eqref{gsscaling} and the fact that the 4d dilaton remains constant in this limit. Using that for a decompactification from 4 to $4+n$ dimensions the species scale is given by
\begin{equation}\label{Lambdasp}
 \Lambda_{\rm sp} = M_{\text{\tiny KK}}^{\frac{n}{n+2}} M_{\rm pl}^{\frac{2}{2+n}}\,,
\end{equation}
we find 
\begin{equation}
 \frac{\Lambda_{\rm sp} }{M_{\rm pl}} \sim \frac{1}{t^{1/2}}\,,
\end{equation}
which is consistent with the asymptotic version \eqref{proposalasymptotic} of \eqref{proposal}. In the mirror dual large complex structure limit the lightest tower of states is given by wrapped D3-branes dual to the D0-branes leading to the same scaling of the species scale \cite{Grimm:2018ohb}.

\subsubsection{Conifold}
Let us now move to the conifold point. As reviewed in the section~\ref{sec:F1review}, in the vicinity of this point $F_1$ diverges as 
\begin{equation}\label{F1coni}
 F_1 \sim -\frac{1}{12} \log |\mu|^2 +\dots \,. 
\end{equation} 
Given the proposed relation between species scale and $F_1$ this would imply that, again, the number of species below the quantum gravity cut-off diverges as we approach the conifold point. However, unlike at the LCS point, at the conifold point there is just a single BPS state that becomes massless; the D3-brane wrapping the $S^3$ that shrinks at the conifold. Hence, the tower of light states contributing to $N_{\rm sp}$ cannot arise from BPS states. Instead, these additional light states have to be non-BPS states. In order for the EFT of the supergravity theory without the new charged hypermultiplet to be still valid we must be sufficiently far away from the conifold point. In other words we are looking for a tower of non-BPS states getting light, at a point where the wrapped conifold branes are still too heavy to change the EFT.

To that end consider the corrections to $F_1$ in \eqref{F1coni}:\footnote{These corrections show up in the real $F_1$ due to the K\"ahler metric term in \eqref{F1general}, but are absent in the topological limit $\bar{\mu} \to 0$, cf.~\cite{Ghoshal:1995wm}.}
\begin{equation}
 F_1= -\frac{1}{12} \log |\mu|^2 +\mathcal{O}( \log [\log |\mu|])\,.
\end{equation}
Given the interpretation of $F_1$ as counting the light states the functional form of the corrections suggests that close to the conifold there are additional light states with mass 
\begin{equation}\label{massconi}
 \frac{M_{\rm light}}{M_s} \sim \frac{1}{(-\log|\mu|)^\gamma}\,, 
\end{equation}
for some $\gamma>0$. In order to correctly compute the species scale associated to this tower of non-BPS states, we need to fix $\gamma$. We therefore need to understand the physical origin of these states which can be traced back using the description of the conifold in terms of the worldsheet CFT \cite{Witten:1995zh,Ooguri:1995wj}. Recall that the mass of the D3-brane becoming light in the vicinity of the conifold is given by 
\begin{equation}
\frac{M_{\rm D3}}{M_s} = \frac{|\mu|}{g_s} \,. 
\end{equation}
In order for the CFT description to be valid and the EFT to continue to be valid, we need the mass of the D3-brane to be above the string scale. To reach the conifold point in the CFT picture we should hence consider the double-scaling limit
\begin{equation}
 \mu, g_s\rightarrow 0 \,,\qquad \frac{|\mu|}{g_s} = \text{const}\, .
\end{equation}
As shown in \cite{Witten:1995zh,Ooguri:1995wj} in this double scaling limit, the Calabi--Yau threefold $X_3$ develops a cigar-shaped throat of length 
\begin{equation}
  \frac{l_{\rm cigar}}{l_s} \sim -\log |\mu|\,, 
\end{equation}
with $l_s$ the string length. Along this cigar the dilaton varies linearly leading to strong coupling as we move down towards the tip of the cigar. The cigar with the linear dilaton profile can be interpreted as the target space for the non-critical $c=1$ string theory at the self-dual radius. The only physical states of the $c=1$ string are localized at the tip of the cigar whereas waves propagating down the throat are BRST trivial. These states can be identified with KK states associated to the extra dimension that becomes large as we approach the conifold point and hence (up to subtleties related to the linear dilaton profile) have mass
\begin{equation}\label{MKKMS}
 \frac{M_{\text{\tiny KK}}}{M_s} \sim \frac{l_s}{l_{\rm cigar}} \sim  \left|\frac{1}{\log |\mu|} \right|\,. 
\end{equation}
The tower of states contributing to $N_{\rm sp}$ in the vicinity of the conifold state can thus be identified with the KK states of the cigar and we can fix $\gamma=1$ in \eqref{massconi}. 

It is now straightforward to compute the species scale associated to these modes using \eqref{Lambdasp}, again with $n=1$. To do that we need the mass of the KK modes in Planck units. For a string compactification to 4d the relation between string and Planck scale is given by 
\begin{equation}
    \frac{M_{\rm pl}^2}{M_s^2} \sim \frac{1}{g_s^2} \cV_{Y_3}\, .
\end{equation}
In the conifold limit, we can expand the volume of the Calabi--Yau threefold as
\begin{equation}
\mathcal{V}_{Y_3} \sim \cV_{\rm rest} \big|\log |\mu|\big|\, .
\end{equation}
Asymptotically the ratio $\cV_{\rm rest}/g_s^2$ is kept constant such that $M_{\rm pl}^2\sim \big|\log|\mu|\big| M_s^2$. Using this and \eqref{MKKMS} the mass of the KK modes on the cigar in Planck units is thus given by 
\begin{equation}
    \frac{M_{\text{\tiny KK}}}{M_{\rm pl}} = \frac{1}{\big|\log |\mu|\big|^{3/2}}\,. 
\end{equation}
Using \eqref{Lambdasp} we then find
\begin{equation}
    \Lambda_{\rm sp}= \left(\frac{M_{\text{\tiny KK}}}{M_{\rm pl}}\right)^{\frac{1}{3}} M_{\rm pl} \sim \left(\frac{1}{\big|\log |\mu|\big|^{3/2}}\right)^{1/3} M_{\rm pl} = \left|\frac{1}{\log |\mu|}\right|^{1/2} M_{\rm pl} \,,
\end{equation}
which is consistent with $N_{\rm sp}\sim \big|\log |\mu|\big| \sim F_1$ in accordance with \eqref{proposalasymptotic}. 

One could have also asked what happens if we are at the conifold point, and not just near it.  In this case we get a massless charged hypermultiplet which modifies the EFT description with the massless mode that we started with.  This would be part of the ground state contribution to $F_1$ that we had to subtract in order to get a finite answer.  In other words, if we were to include the wrapped conifold brane which is massless, we would get a different $F_1$ which would have no divergence at the conifold point, as the charged massless field would be already accounted for by the degrees of freedom of the EFT.  It is natural to expect that the $\log(|\mu|)$ term in such a case would lead to the shift of the cutoff and serves as an additive constant to $F_1$ as already discussed.  It would be interesting to study this further, which we will not attempt in this paper.

\subsubsection{K-point}\label{sec:Kpoint}
There is an additional class of singularities that we may consider corresponding to infinite distance limits in the moduli space without maximally unipotent monodromy. In one-parameter moduli spaces, such points correspond to so-called `K-points'. Let us choose coordinates on the moduli space such that the K-point is located at $\psi=0$.\footnote{Here we assume that any semi-simple part of the monodromy has been removed by sending $\psi \to \psi^k$.} Then in the vicinity of $\psi=0$ to leading order we have 
\begin{equation}
    F_1 = - \log |\psi|^2 + \dots\,. 
\end{equation}
To compare this with the species scale, we need to have information about the leading tower of states becoming light at the K-point. To that end, let us consider the analogue of K-points in higher-dimensional moduli spaces which arise for instance if the mirror $X_3$ of $Y_3$ is K3-fibered. Therefore, let $X_3$ be a CY3-fold that is a K3-fibration over $\mathbb{P}^1$. We can expand the K\"ahler form as 
\begin{equation}
    J_0 = v^0 J_0 + \sum_{a=1}^{h^{1,1}-1} v^a J_a \,,
\end{equation}
where $v^0$ is the volume of the base $\mathbb{P}^1$ and $v^a$ are the volumes of fibral curves. The analogue of the K-point now corresponds to the limit 
\begin{equation}\label{emerstringlimit}
    v^0 \rightarrow \infty, \qquad v^a =\text{const}, \quad \forall a\,. 
\end{equation}
In \cite{Lee:2019oct}, such a limit in the Type IIA vector multiplet moduli space was identified as an emergent string limit in which an NS5-brane wrapped on the generic K3-fiber, which is dual to the critical heterotic string on $K3\times T^2$, becomes tensionless and weakly coupled. In Planck units the tension of this string is simply given by 
\begin{equation}\label{Thet}
    \frac{T_{\rm het}}{M_{\rm pl}^2} \sim \frac{1}{v^0}\,. 
\end{equation}
As pointed out in \cite{Lee:2019oct}, the tower of string excitations is also the leading tower of light states therefore setting the species scale. Due to the exponential degeneracy of string states the species scale is in fact of the order of string scale (up to a logarithmic rescaling \cite{Marchesano:2022axe,Castellano:2022bvr})
\begin{equation}\label{speciesscaleemstring}
    \frac{\Lambda_{\rm sp}^2}{M_{\rm pl}^2} \sim \frac{1}{v^0} \,.
\end{equation}
Again, we can compare this with $F_1$ in the limit \eqref{emerstringlimit} which according to \eqref{asymptLCS} scales as 
\begin{equation}\label{F1emstring}
    F_1 \rightarrow \frac{1}{12} c_{2,i} \text{Re}\,t^i \sim \frac{1}{12} c_{2,0} v^0 = 2 v^0\,,
\end{equation}
where $\text{Re}\,t^0=v^0\gg v^a = \text{Re}\,t^a$ and we used $c_{2,0}=24$ since $J_0$ is the class of the generic K3-fiber. Comparing \eqref{speciesscaleemstring} and \eqref{F1emstring}, we again find as expected
\begin{equation}
    \frac{\Lambda_{\rm sp}}{M_{\rm pl}}\sim \frac{1}{\sqrt{F_1}}\, .
\end{equation}
Let us briefly comment on the case that instead of a K3-fibration, $X_3$ is fibered by an Abelian surface. Again, $J_0$ is the class of the generic surface fiber, but we have 
\begin{equation}
    c_{2,0} = 0\,, 
\end{equation}
implying that $F_1$ does not diverge in the limit \eqref{emerstringlimit}. However, an NS5-brane on the generic fiber gives rise to a critical Type II string \cite{Lee:2019oct} with tension scaling as in \eqref{Thet} such that the species scale scales again like
\begin{equation}
    \frac{\Lambda_{\rm sp}}{M_{\rm pl}}\sim \frac{1}{\sqrt{v^0}}\,. 
\end{equation}
To explain this mismatch, let us recall that $F_1$ only computes an index such that there can occur cancellations between fermions and bosons even though both contribute to $N_{\rm sp}$ (see our discussion below \eqref{NspF1}). A systematical cancellation arises in case of higher supersymmetry. In the case of a $T^4$-fibered CY3-fold the supersymmetry is not enhanced to $\cN=4$, but there is an $\cN=4$ sub-sector. This sub-sector precisely corresponds to the NS5-brane on the Abelian fiber which can be identified with a Type II string with enhanced $(4,4)$ world-sheet supersymmetry. In the large base limit \eqref{emerstringlimit} all supersymmetry-breaking effects get diluted. Accordingly also the spectrum of light states resembles a $\cN=4$ spectrum for which, however, the contribution to $F_1$ vanishes identically. Due to this systematic cancellation in this case we cannot identify $F_1$ with the species scale in accordance with our discussion below \eqref{NspF1}.

\section{The species scale and the desert in explicit models}\label{sec:ex}
In this section we illustrate the main idea laid out in the previous section in some examples. Focusing again on the vector multiplet sector, we explicitly identify the point where $F_1$ is minimized, i.e.~where our proposed species scale \eqref{proposal} is maximized. We also compare these findings with the behavior of the masses of some of the BPS states, that is, the ones that vanish at some singularities in the moduli space.

To this end, let us briefly recall some facts about BPS states and their masses in $\mathcal{N}=2$ supergravity theories. We focus on Type IIB Calabi--Yau compactifications, in which case the BPS states arise from D3-branes wrapping special Lagrangian three-cycles in the geometry. Let us denote by $q \in H^3(Y_3, \mathbb{Z})$ the three-forms Poincar\'e dual to the three-cycles. The central charge of the corresponding BPS state is then given by 
\begin{equation}
    Z(q) = e^{K/2} \int_{Y_3}  q \wedge \Omega\, ,
\end{equation}
with $\Omega$ the $(3,0)$-form of the Calabi--Yau manifold. The mass of the BPS states is then simply given by $M(q)=|Z(q)|$. We can compute these masses and their complex structure moduli dependence by expanding $\Omega$ in terms of so-called period integrals as
\begin{equation}
\Omega(x) = \Pi^I(x) \gamma_I\, ,
\end{equation}
with an integral three-form basis $\gamma_I \in H^3(Y_3, \mathbb{Z})$ ($I=1,\ldots, 2h^{2,1}+2$). As we elaborate upon in appendix \ref{app:periods}, we can evaluate these periods explicitly and thereby describe the dependence of the BPS masses on the complex structure moduli. In the following we use these methods to characterize the behavior of both, our species scale proposal \eqref{proposal}
and the mass gap for the BPS states.

\subsection{\texorpdfstring{$T^2$}{T2}}
\begin{figure}[!ht]
\begin{center}
\includegraphics[width=10cm]{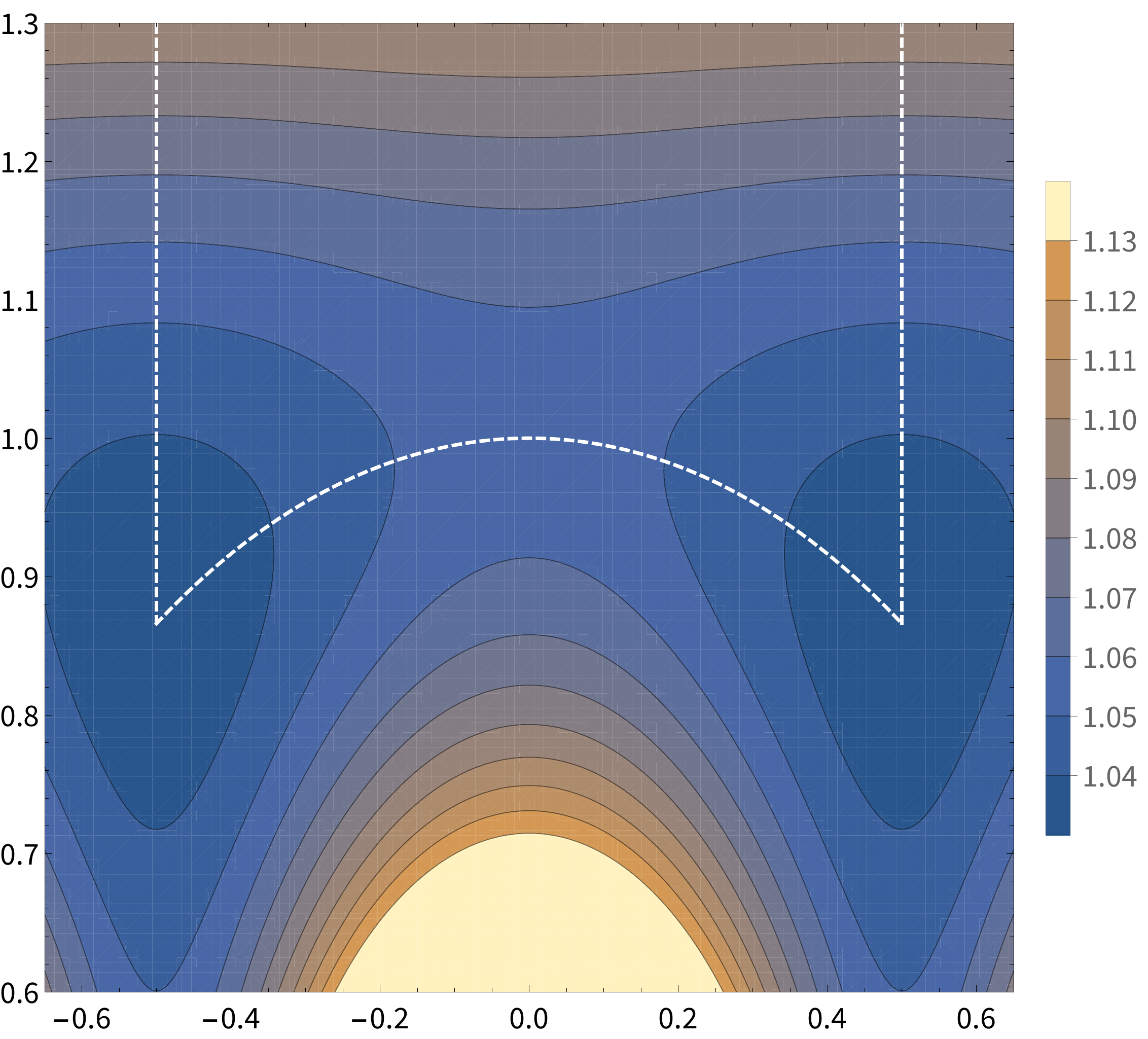}
\end{center}
\begin{picture}(0,0)\vspace*{-1.2cm}
\put(228,22){$\tau_1$}
\put(88,155){$\tau_2$}
\end{picture}\vspace*{-0.8cm}
\caption{\label{fig:F1T2} Contour plot of the genus one free energy $F_1$ in the upper half plane for $T^2$. The fundamental domain has been indicated by a dashed, white line. The minimum of $F_1$ is located precisely at the cusp $\tau=e^{2\pi i/3}$ (and $\tau=e^{\pi i /6}$). }
\end{figure}
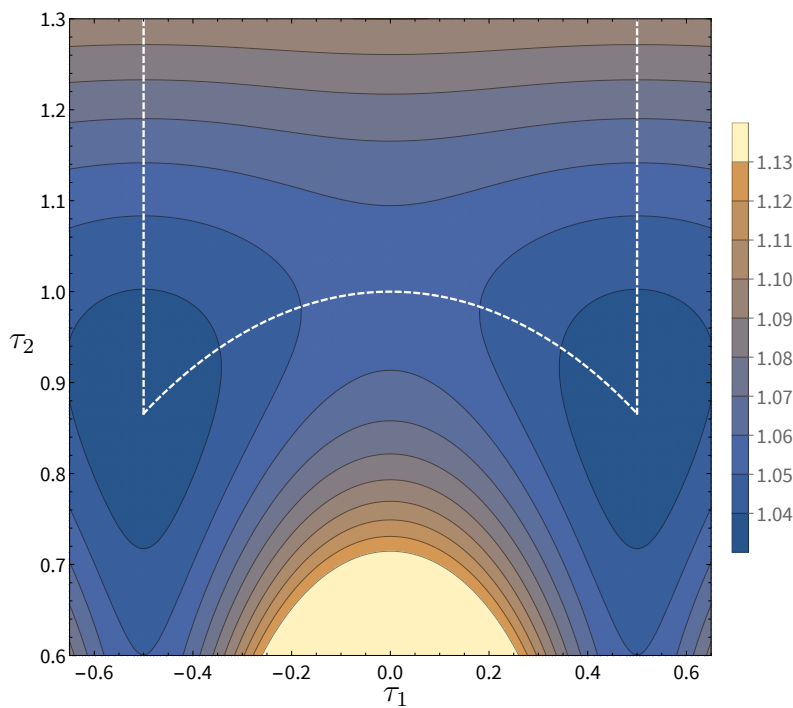

To start off, we test our idea in the simple example where the target space of the topological string is $T^2$.  This can be viewed as part of a CY 3-fold (for instance $K3\times T^2$). The complex structure moduli space of $T^2$ is the fundamental domain of $\text{SL}(2,\mathbb{Z})$, and arises for instance as the axio-dilaton field space of Type IIB string theory. In the latter setup the BPS mass gap has readily been studied in \cite{Long:2021jlv}, with the cusp at the third root of unity as the desert point. Here we compare these findings with the behavior of our proposed species scale. For this setting the genus one free energy of the topological string has been derived in \cite{Bershadsky:1993ta}
\begin{equation}\label{eq:F1T2}
    F_1 = -\log \tau_2 |\eta^2(\tau)|^2\, ,
\end{equation}
where $\tau=\tau_1+i \tau_2$ is the complex structure parameter, and $\eta(\tau)$ is the Dedekind eta function. Since $F_1$ is inversely related to the species scale, the minimum of $F_1$ corresponds to its desert point. By evaluating \eqref{eq:F1T2} explicitly on the upper half plane, we find that it is minimized at the cusp of the fundamental domain $\tau = \exp(2\pi i /3)=-\frac{1}{2}+\frac{\sqrt{3}}{2}$. For completeness, let us record its numerical value at this point as
\begin{equation}
    F_1 \big|_{\tau = \exp(2\pi i/3)} = -\log\left[\frac{3 \Gamma(\frac{1}{3})^{1/6}}{32 \pi^4}\right] \simeq 1.03352 \, ,
\end{equation}
although we note that, since $F_1$ is only defined up to an additive constant, this number does not carry much information. In figure~\ref{fig:F1T2}, we provided a contour plot to illustrate the behavior of $F_1$ on the upper half plane.

Next, we check whether this minimum of $F_1$ -- the maximum of our species scale -- matches with the maximum of the BPS mass gap. To this end, we turn to \cite{Long:2021jlv} where the tensions of $(p,q)$-strings of ten-dimensional Type IIB string theory were studied. These tensions are given by
\begin{equation}
T_{(p,q)} \sim |p-q\tau|/\tau_2\, .
\end{equation}
Finding the mass gap then amounts to minimizing over all tensions $T_{(p,q)}$ with $p,q \in \mathbb{Z}$, which was found to be equivalent to the classic problem of sphere packing \cite{chang2010simple}. The maximum gap for the tensions then, indeed, corresponds to the cusp at $\tau = \exp(2\pi i/3)$. The lightest strings at this point are the F1-string, the D1-brane, and the $(1,1)$-string -- all with equal tensions  $T_{(1,0)}=T_{(0,1)}=T_{(1,1)}$. Moreover, these strings are mapped into each other by the orbifold $\mathbb{Z}_3$ symmetry associated to this point.

\subsection{The mirror quintic \texorpdfstring{$X_5(1^5)$}{X5}}\label{sec:X5}
For our next example we look at the mirror quintic. We parametrize the moduli space by the coordinate $x$ in which the large complex structure point lies at $x=0$, the conifold point at $x=5^{-5}$ and the Landau-Ginzburg point at $x=\infty$. The relation to the standard coordinate for the Landau-Ginzburg point used in \cite{Candelas:1990rm} is given by $x = 1/(5 \psi)^5$, where the quintic is defined as
\begin{equation*}
    X_{5}=\left\{(z_0,z_1,\dots,z_4)\in\mathbb{P}^4\left\vert \sum_{i=0}^4z_i^5-5\psi z_0z_1z_2z_3z_4=0\right.\right\}\,.
\end{equation*}
For the quintic the genus one free energy is given by
\begin{equation}\label{eq:F1quintic}
F_1 = \frac{31}{3}K-\frac{1}{2}\log[ G_{x \bar{x}}]+\log \left|x^{-\frac{31}{12}} (1-5^5 x)^{-1/12}\right|^2\, ,
\end{equation}
where the holomorphic ambiguity $f(x)=x^{-\frac{31}{12}} (1-5^5 x)^{-1/12}$ in \eqref{F1general} has been fixed by requiring the correct asymptotic behaviors at the conifold point \eqref{asymptconi} and the LCS point \eqref{asymptLCS} with $c_2=50$. Note that the holomorphic ambiguity looks slightly different compared to \cite{Bershadsky:1993ta} since we work in the LCS coordinate $x$ instead of the Landau-Ginzburg coordinate $\psi$.

\begin{figure}[!ht]
\vspace*{2pt}
\centering
\begin{subfigure}{0.49\textwidth}
\centering
\includegraphics[width=230pt]{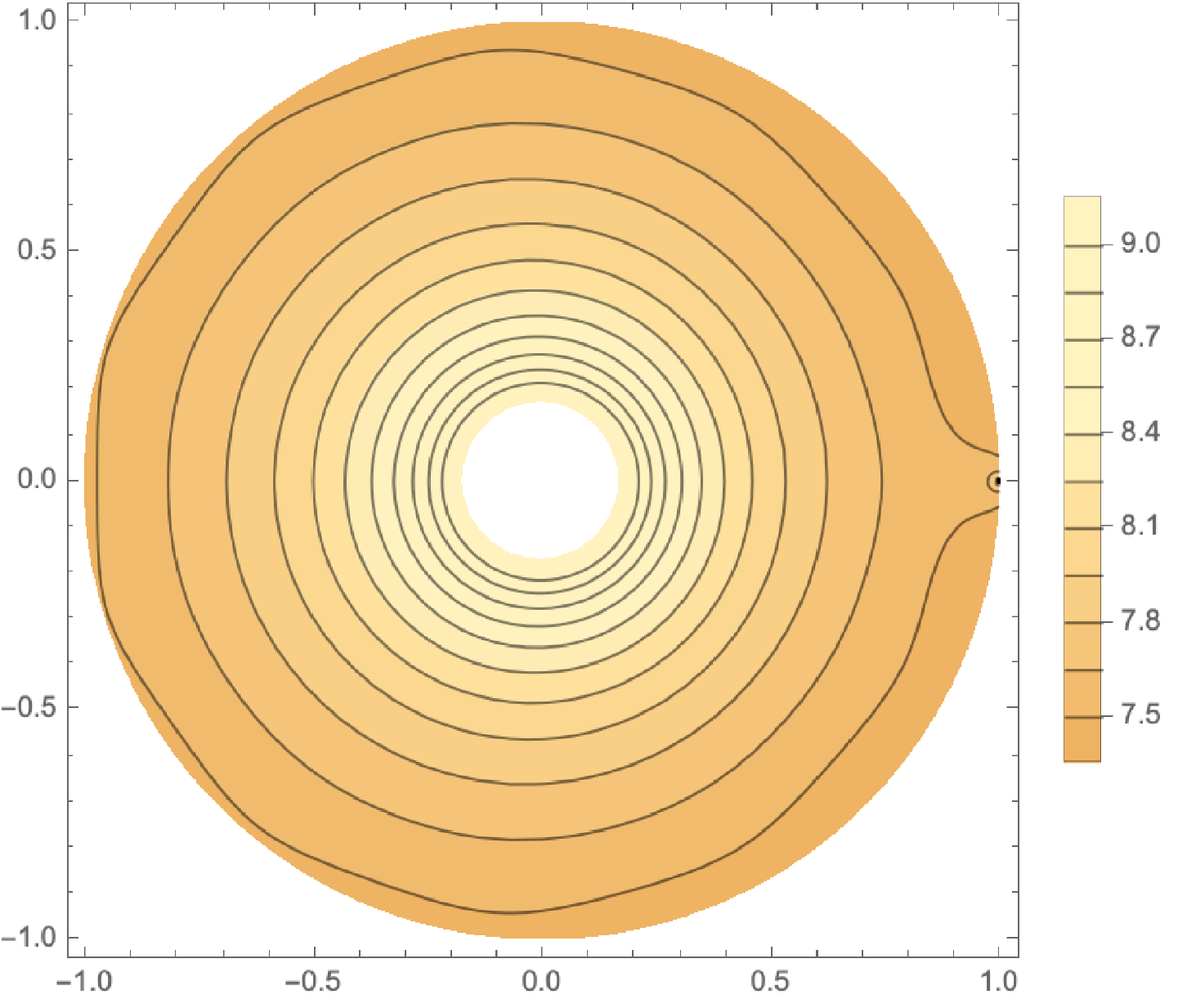}%
\vspace{-20pt}\caption{LCS point in  $x_{\rm LCS} =5^{5}x $. \label{fig:LCS}}
\end{subfigure}
\hspace{2pt}
\begin{subfigure}{0.49\textwidth}
\centering
\includegraphics[width=230pt]{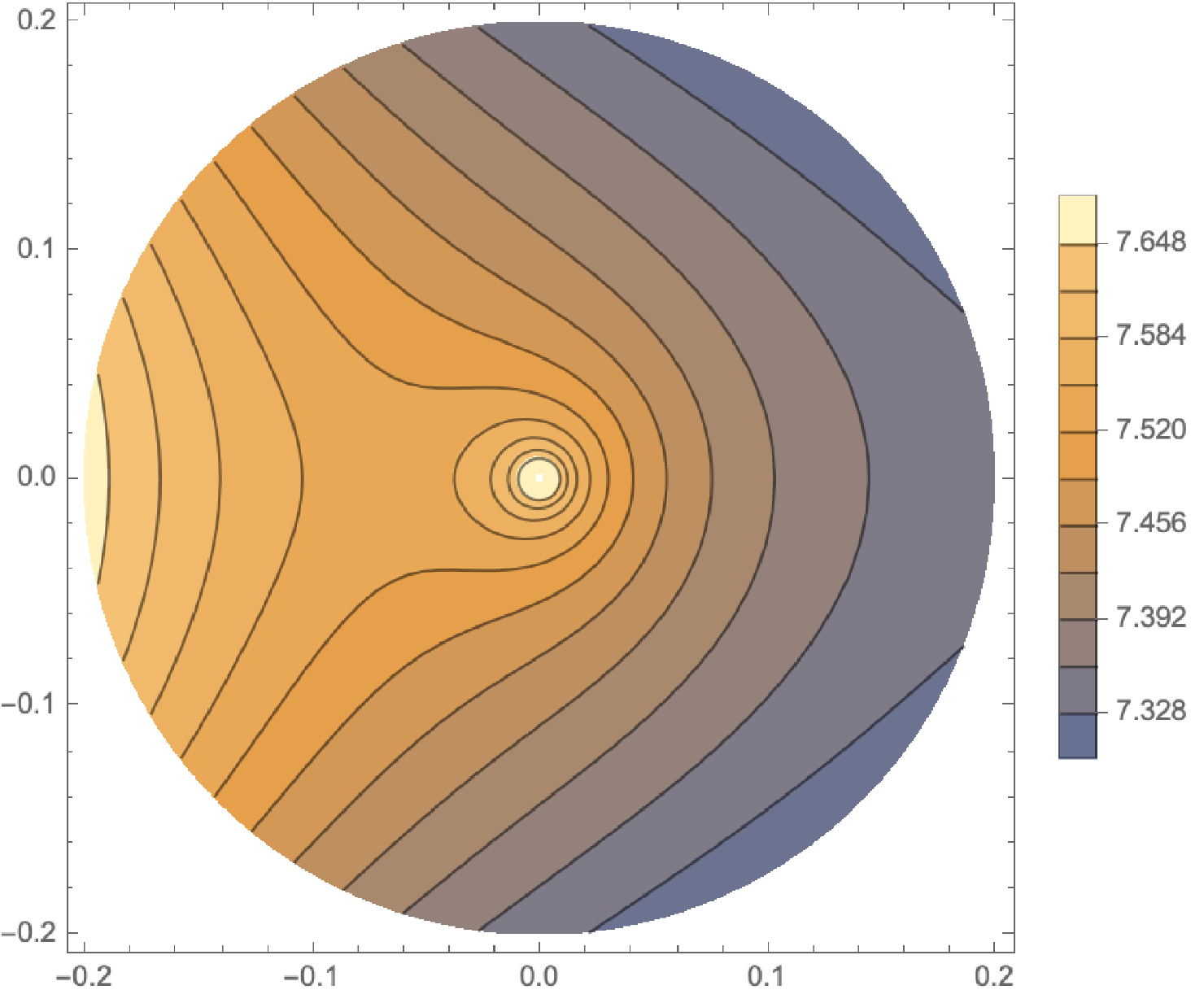}%
\vspace*{-20pt}\caption{Conifold point in  $\mu= 5^{5}(x-5^{-5})$.\label{fig:quinticconifold}}
\end{subfigure}

\begin{subfigure}{0.49\textwidth}
\centering
\vspace*{10pt}\includegraphics[width=230pt]{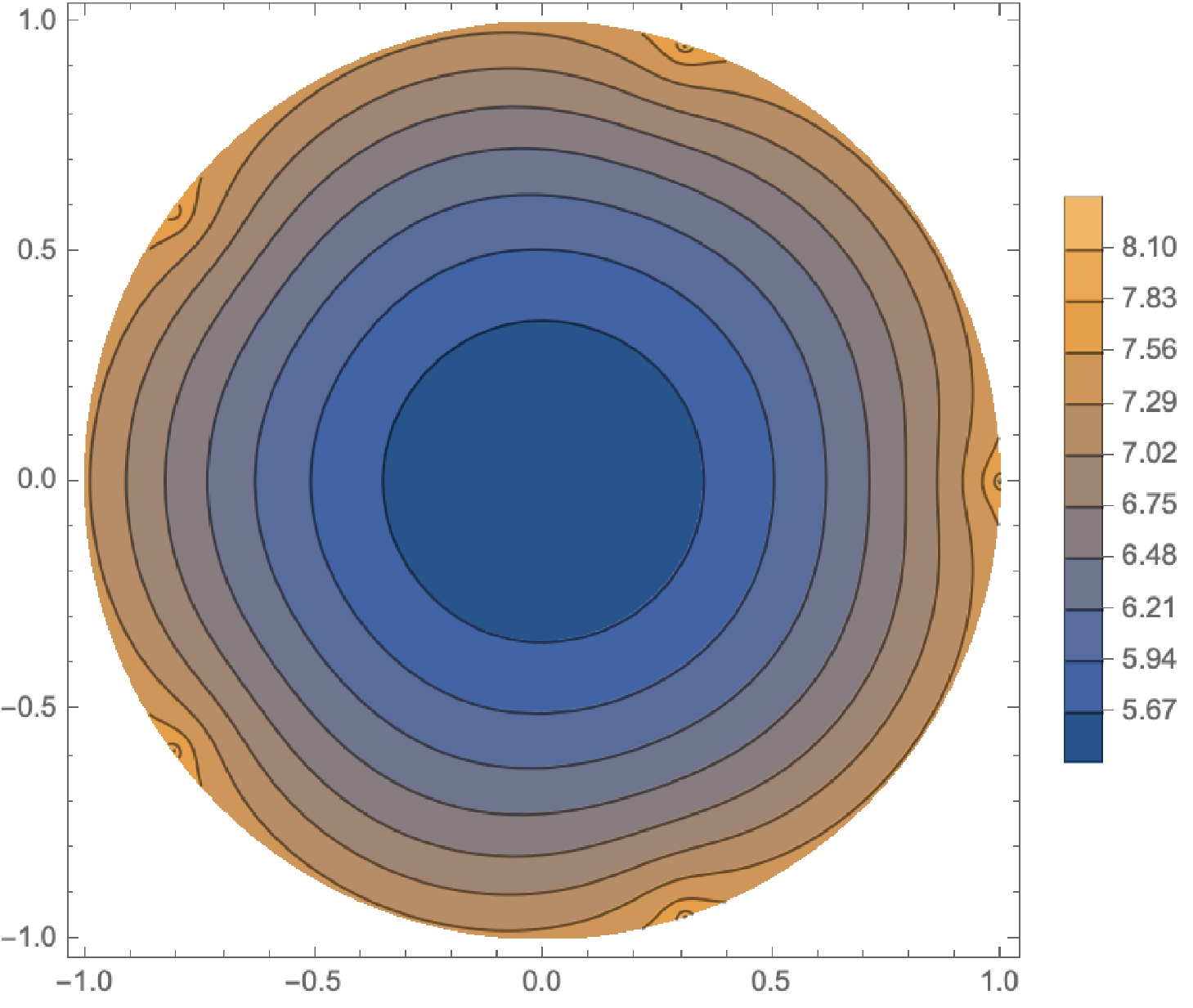}%
\vspace*{-20pt}\caption{LG-point in $\psi= \frac{1}{5}x^{-1/5}$.\label{fig:quinticLG}}
\end{subfigure}
\caption{Contour plots of $F_1$ in the three patches in the moduli space of the quintic: near the LCS point, the conifold point, and the Landau-Ginzburg point. The genus one free energy $F_1$ has an absolute minimum at the Landau-Ginzburg point $x=\infty$. For the LCS patch in fig.~\ref{fig:LCS} it decreases radially as we move away from the singularity, with a rotational inhomogeneity due to the conifold point at $x_{\rm LCS}=1$. For the conifold patch in fig.~\ref{fig:quinticconifold} we see a singularity at $\mu=0$ where $F_1$ diverges. For the LG-point patch in fig.~\ref{fig:quinticLG} we see that $F_1$ increases as we move out radially, with a rotational inhomogeneity due to the conifold points at the fifth roots of unity. \label{fig:F1quintic}}
\end{figure}

We now want to determine the point at which our species scale is maximized, i.e.~where \eqref{eq:F1quintic} is minimized in moduli space. Recall from section~\ref{sec:F1review} that $F_1$ diverges at the LCS point and conifold point. At the LG-point, however, $F_1$ remains finite, so this is a natural candidate for the desert point. Another possibility would be the region in between the LCS point and the conifold point, but here $F_1$ turns out to take larger values than at the LG-point. By scanning $F_1$ explicitly over the moduli space we verify that the global minimum of $F_1$ indeed lies at the LG-point. To illustrate these findings we have included three contour plots in figure~\ref{fig:F1quintic}, for the three different patches in the moduli space.

We now compare this minimum of $F_1$ with the location of the maximal BPS mass gap in the 4d theory. Let us first characterize the states we want to consider. From the Type IIB perspective these BPS states arise from D3-branes wrapped on special Lagrangian three-cycles of the Calabi--Yau manifold. The light states in the theory typically correspond to D3-branes that become massless at some singularity in moduli space. In order to sketch a more intuitive picture for these light states, let us use the language of the mirror Type IIA setup. For the large complex structure point we then know that a tower dual to D2-D0 branes is becoming massless, while for the conifold point a single state dual to the D6-brane becomes massless. From the perspective of the LG point these are the light BPS states in one corner of the moduli space, and by using its monodromy we can generate the light states in the other four corner.

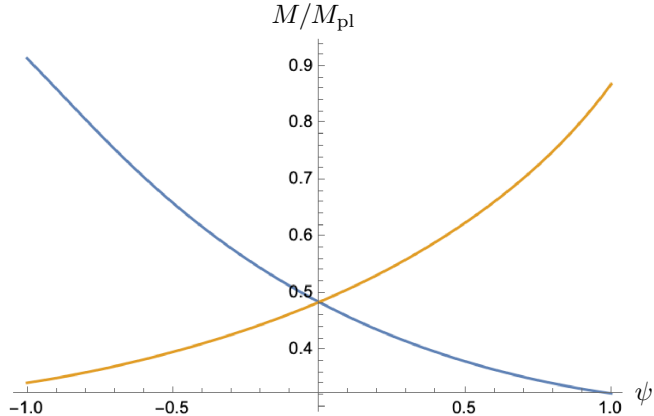
\begin{figure}[!ht]
\centering
\vspace*{10pt}
\includegraphics[width=230pt]{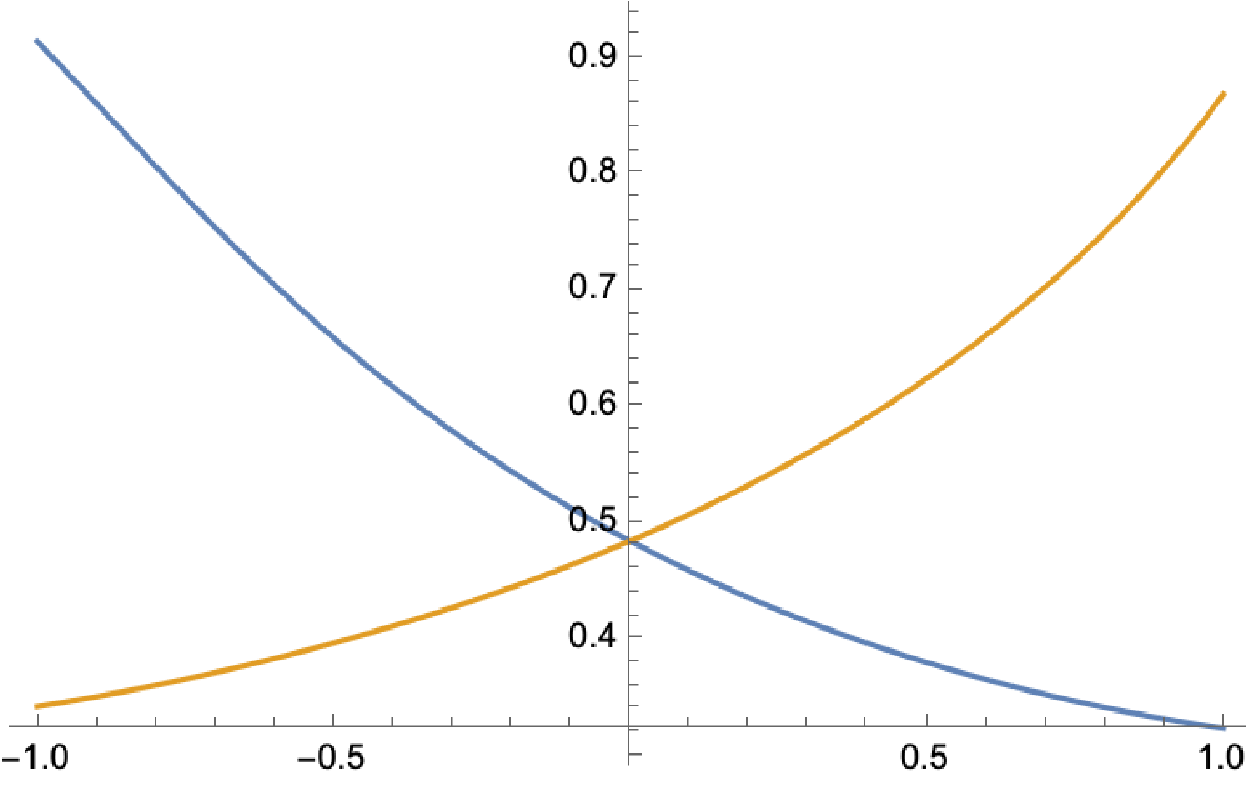}%
\begin{picture}(0,0)
\put(-132,150){\footnotesize$M/M_{\rm pl}$}
\put(4,8){\footnotesize$\psi$}
\end{picture}
\caption{Plot of the masses of the D2-brane (blue) and its copy under two LG-point monodromies (yellow) along the real line in the complex $\psi=\frac{1}{5} x^{-1/5}$ plane. \label{fig:MD2quintic}}
\end{figure}

Next we want to investigate how the masses of these light BPS states vary over the moduli space. For instance, say we move from one LCS point to another copy under the Landau-Ginzburg monodromy. Then the light states of this first LCS point will become more heavy, while the states associated to the second LCS point will become lighter. The task of finding the BPS desert point thus amounts to determining where the masses of these states cross with a maximal gap. A natural candidate for this point is the Landau-Ginzburg point itself due to its orbifold $\mathbb{Z}_5$ symmetry. Evaluating the masses of the light BPS states at the LG-point we find that the D2-brane state (and its LG-point monodromy copies) are the lightest states. For completeness, let us write down the values of the masses of the lightest states explicitly
\begin{equation}
    \begin{aligned}
    \frac{M_{\rm D2}}{M_{\rm pl}} &= \frac{\sqrt[4]{\frac{1}{2} \left(5-\sqrt{5}\right)}}{\sqrt{5}} \simeq 0.484886\, , \qquad \frac{M_{\rm D0}}{M_{\rm pl}} = \frac{1}{\sqrt[4]{5+2 \sqrt{5}}} \simeq 0.570017 \, , \\
    \frac{M_{\rm D6}}{M_{\rm pl}} &= \sqrt[4]{\frac{1}{2} \left(25-11 \sqrt{5}\right)} \simeq 0.670095\, .
    \end{aligned}
\end{equation}
Furthermore, from every angle of approach to the LG-point there is always one of the D2-brane states decreasing in mass while another of its LG-point monodromy copies increases. This means that the LG-point indeed maximizes the mass gap of the BPS states, since at other points there will always be a lighter state present. This point has been illustrated in figure~\ref{fig:MD2quintic} for the slice along the real line. 

\subsection{The mirror of \texorpdfstring{$X_{4,2}(1^6)$}{X42}}\label{sec:X42}
In the previous two examples we considered moduli spaces with an orbifold point. As we have seen, these singularities are a natural desert point for our proposed species scale, but they are not present in every moduli space. To this end, we take now an example that does not feature orbifold points: the mirror of $X_{4,2}$ (the intersection of a quartic and a quadric) in $\mathbb{P}_{1,1,1,1,1,1}$. The one-parameter variation of Hodge structure of this Calabi--Yau threefold corresponds to the hypergeometric family AESZ 6 in the database \cite{AESZ}.

The moduli space contains three singularities: a large complex structure point at $x=0$, a conifold point at $x=1$ and another conifold point at $x=\infty$. The monodromy around the second conifold contains a semi-simple part of order four (the local periods have exponents $(\frac{1}{4}, \frac{1}{2}, \frac{1}{2}, \frac{3}{4})$). Moreover, from the monodromy data in \eqref{eq:AESZ6modG} we read off that the $S^3$ that shrinks at the conifold point is quotiented by a finite group of order $|G|=2$.

For this geometry the genus one free energy of the topological string is given by
\begin{equation}
F_1 = \frac{28}{3} K -\frac{1}{2} \log[G_{x\bar{x}}]+ \log |x^{-\frac{17}{6}}(1-1024x)^{-1/12}|^2\, .
\end{equation}
The holomorphic ambiguity $f(x)=x^{-\frac{17}{6}}(1-1024x)^{-1/12}$ has been fixed by requiring the correct asymptotics at the conifold point at $x=1/1024$ (see \eqref{asymptconi}) and the LCS point \eqref{asymptLCS} with $c_2=56$. As a consistency check we can consider the asymptotics at the second conifold at $x=\infty$: Here $F_1$ diverges as $-\frac{2}{12} \log |\psi|^2$, with $\psi$ the local coordinate given by $x = 1024 \psi^4 $ without any finite order monodromies, which is precisely the required coefficient for the quotient group $|G|=2$.

\begin{figure}[!ht]
\vspace*{2pt}
\centering
\begin{subfigure}{0.49\textwidth}
\centering
\includegraphics[width=230pt]{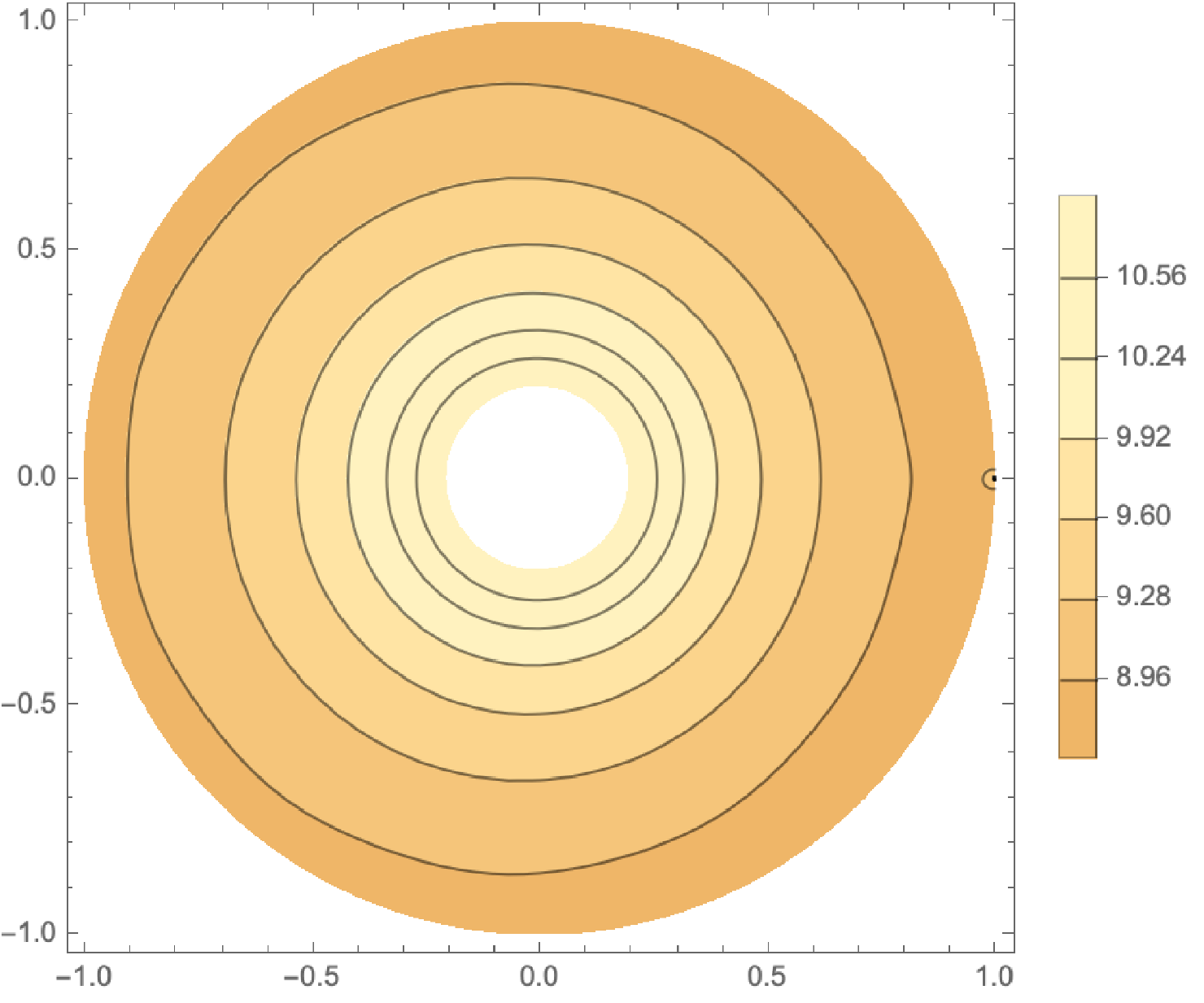}%
\vspace{-20pt}\caption{LCS point in  $x_{\rm LCS} =1024 x $. \label{fig:6LCS}}
\end{subfigure}
\hspace{2pt}
\begin{subfigure}{0.49\textwidth}
\centering
\includegraphics[width=230pt]{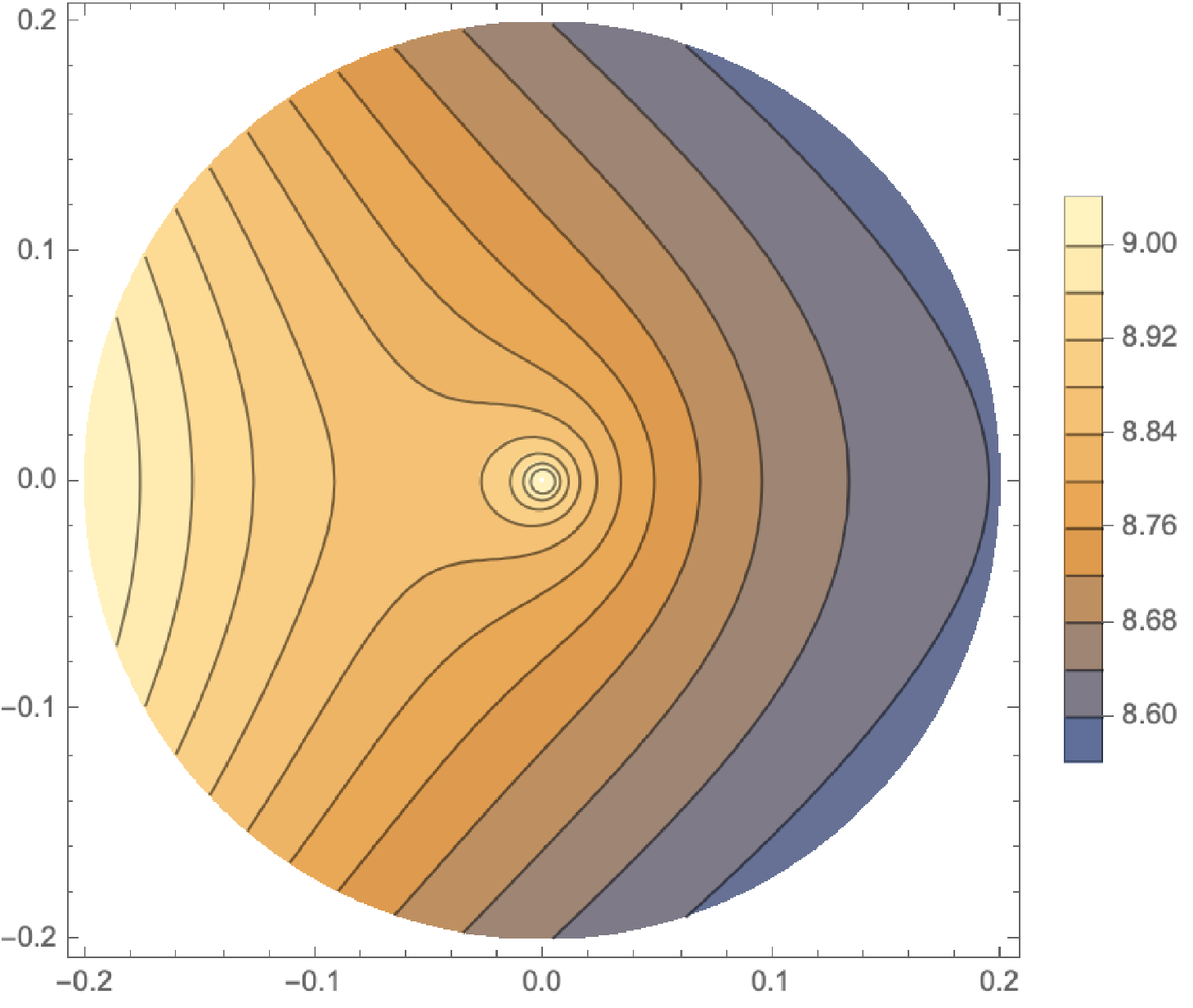}%
\vspace*{-20pt}\caption{Conifold point in  $\mu= 1024(x-1/1024)$.\label{fig:6conifold}}
\end{subfigure}

\begin{subfigure}{0.49\textwidth}
\centering
\vspace*{10pt}\includegraphics[width=230pt]{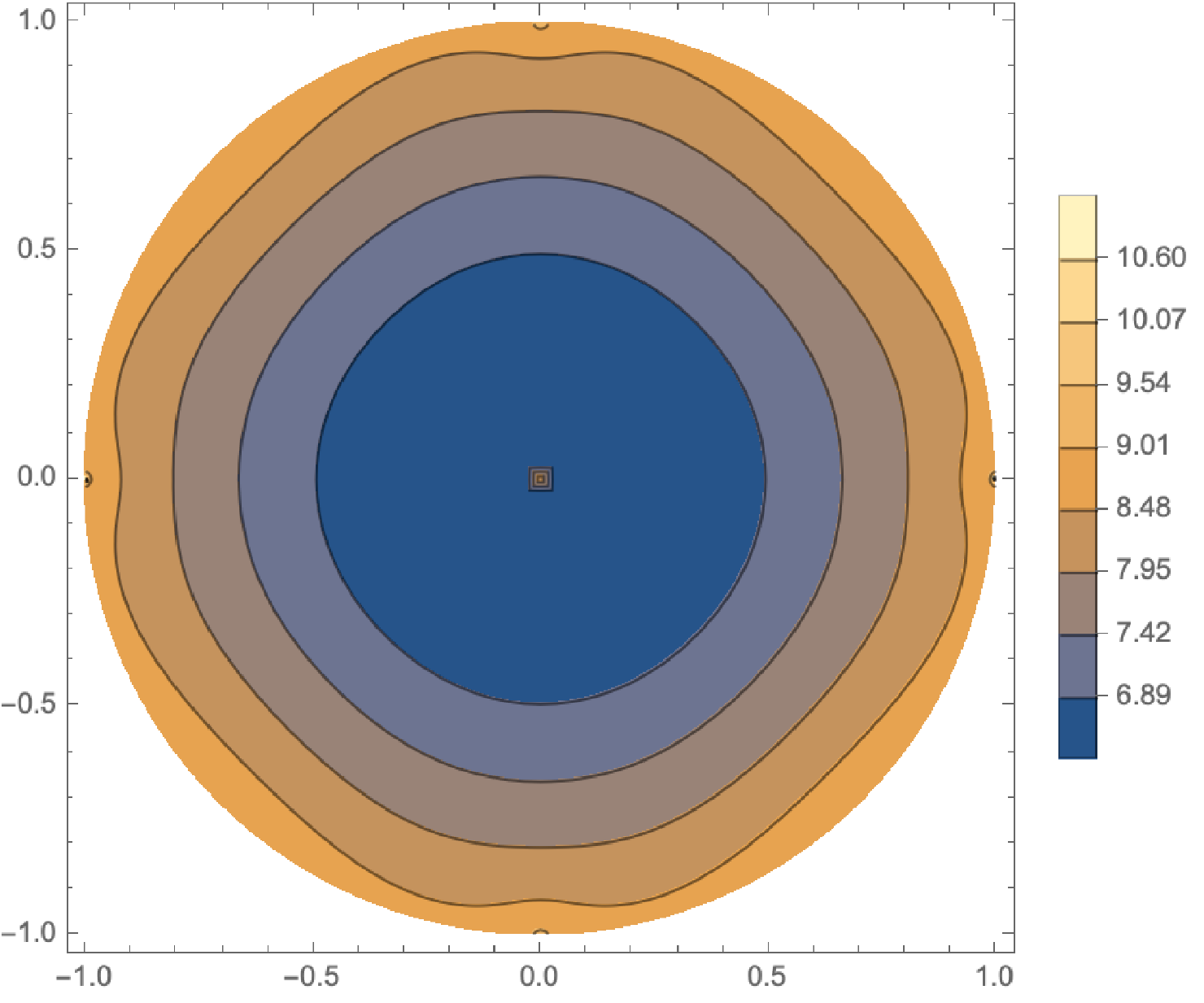}%
\vspace*{-20pt}\caption{Second conifold in $\psi= \frac{1}{4\sqrt{2}}x^{-1/4}$.\label{fig:6infconifold}}
\end{subfigure}
\hspace{2pt}
\begin{subfigure}{0.49\textwidth}
\centering
\vspace*{10pt}\includegraphics[width=230pt]{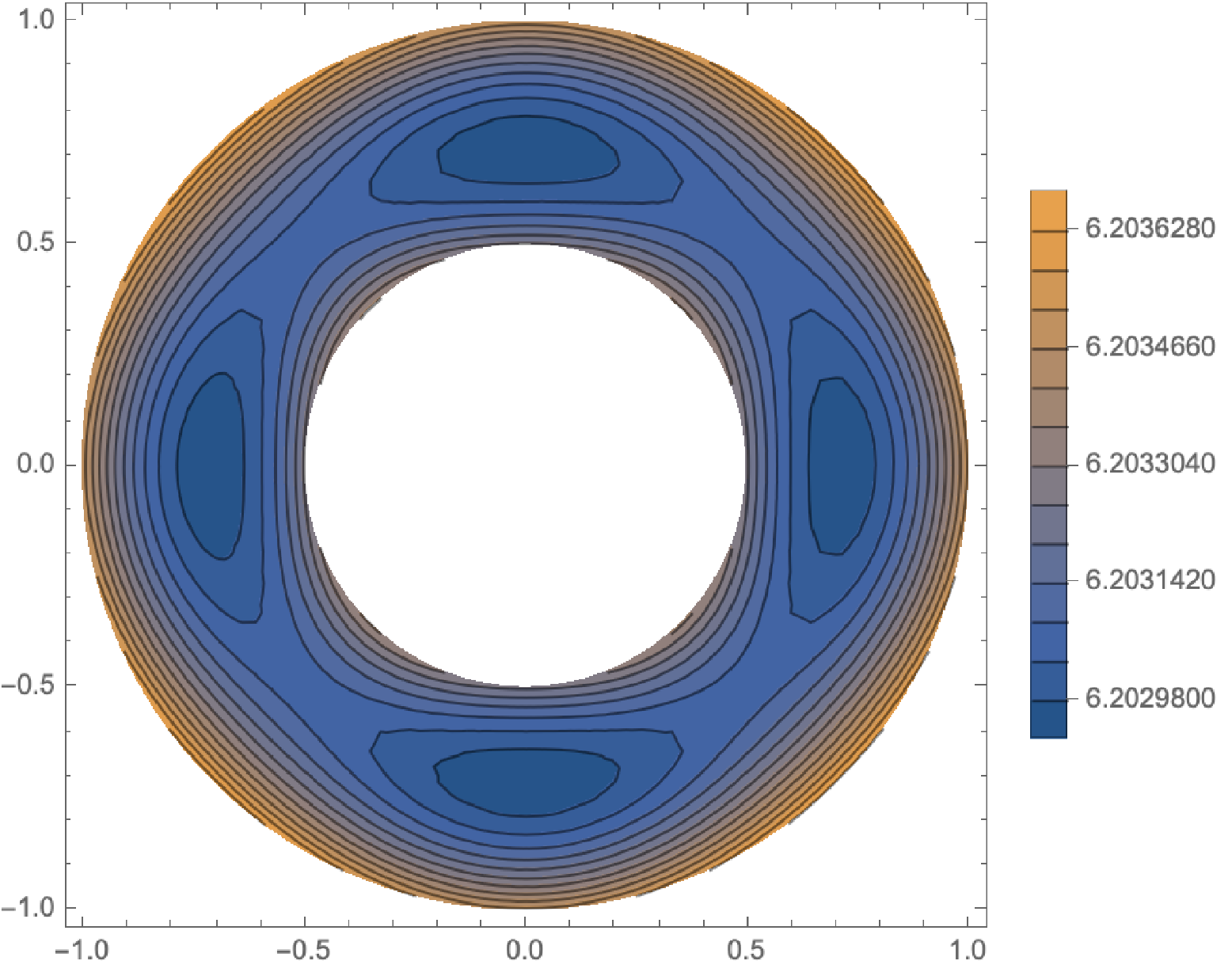}%
\vspace*{-20pt}\caption{Second conifold in $\psi_{\rm res}=\frac{ \psi}{40 |\psi|}(|\psi|+4) $.\label{fig:6infconifoldzoom}}
\end{subfigure}
\caption{Contour plots of $F_1$ in the moduli space of $X_{4,2}$ in $\mathbb{P}_{1,1,1,1,1,1}$. For the LCS patch in fig.~\ref{fig:6LCS} it decreases radially as we move away from the singularity, with a rotational inhomogeneity due to the conifold point at $x_{\rm LCS}=1$. For the conifold patch in fig.~\ref{fig:6conifold} we see a singularity at $\mu=0$ where $F_1$ diverges. For the conifold point at $x=\infty$ we included two figures: in \ref{fig:6infconifold} we see a band around the conifold point where $F_1$ takes the lowest value; in \ref{fig:6infconifoldzoom} we rescale to coordinate to zoom in, from which it becomes clear that $F_1$ is minimized precisely on the straight lines between the conifold point at $\psi=0$ and the conifold points at the fourth roots of unity. \label{fig:F1AESZ6}}
\end{figure}

Next we investigate where $F_1$ is minimized, i.e.~the location of the desert for our species scale proposal. Since $F_1$ diverges at all singular points in moduli space, we have to look elsewhere for this minimum. We perform contour plots over the moduli space in the patches around the singularities, which have been included in figure~\ref{fig:F1AESZ6}. From these contour plots we read off that $F_1$ takes the lowest values on a circle around the conifold point at infinity. Zooming in closer on this circle, we see that there are four minima for $F_1$, related by the finite order part of the monodromy around this singularity. These minima lie precisely on the lines between the conifold point at infinity and the other conifold points at the fourth roots of unity. For completeness, let us give the precise location in the Picard--Fuchs coordinate
\begin{equation}
x_{\rm desert} \simeq  5.06827\, .
\end{equation}
At this point the periods do not satisfy any special algebraic relations compared to other points in its vicinity, so we do not write them down here.

\begin{figure}[!ht]
\centering
\vspace*{10pt}
\includegraphics[width=230pt]{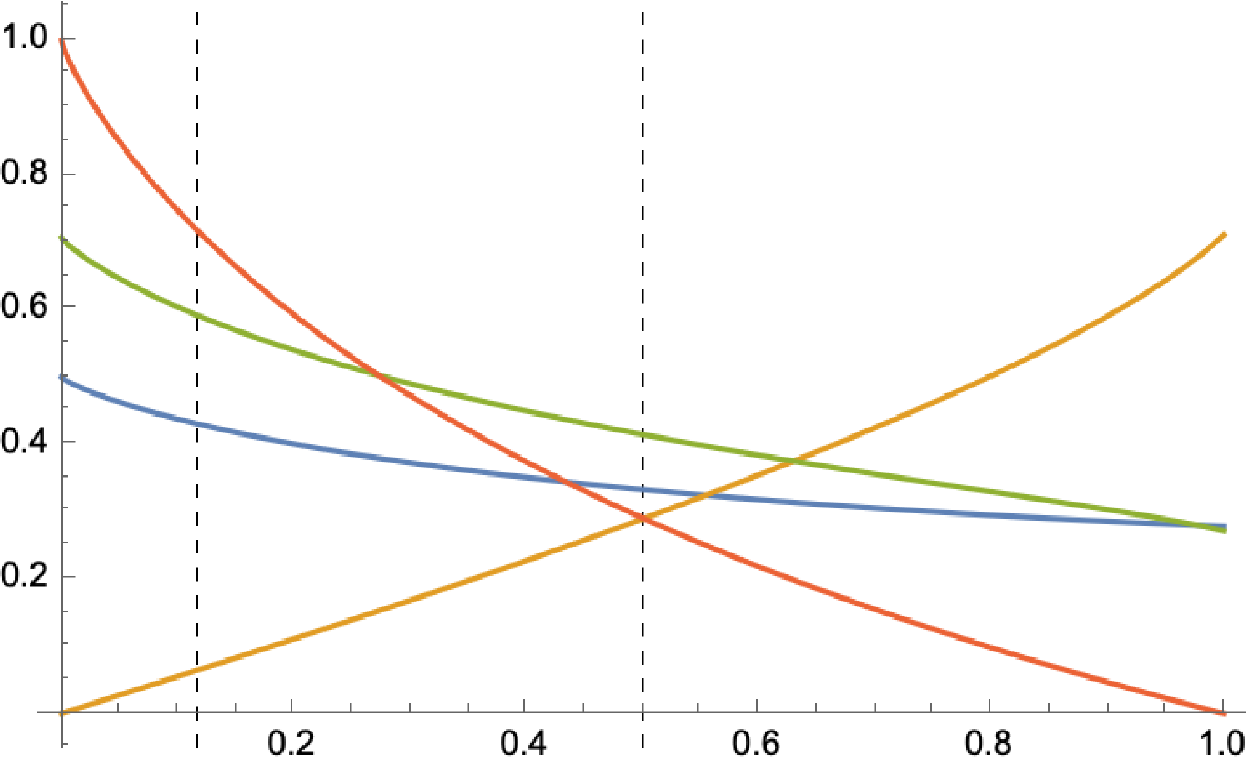}%
\begin{picture}(0,0)
\put(-235,150){\footnotesize$M/M_{\rm pl}$}
\put(4,8){\footnotesize$\psi$}
\put(-213,-8){\footnotesize min.~$F_1$}
\put(-137,-8){\footnotesize max.~$\Lambda_{\rm BPS}$}
\end{picture}\vspace{4mm}
\caption{Plot of the masses of the 2D6+D4+4D0-brane (yellow), D2-brane (blue), D6-brane (red) and the D0-brane (green) in the coordinate $\psi$ defined by $x= 1/(1024 \psi^{4})$. In this coordinate the conifold at $x=\infty$ is located at $\psi=0$ and the conifold at $x=1/1024$ at $\psi=1$. The minimum of $F_1$ is located at $\psi=0.117817$, while the maximum gap of the BPS states in the above plot is at $\psi=0.501072$, as has been indicated by dashed lines.\label{fig:6masses}}
\end{figure}
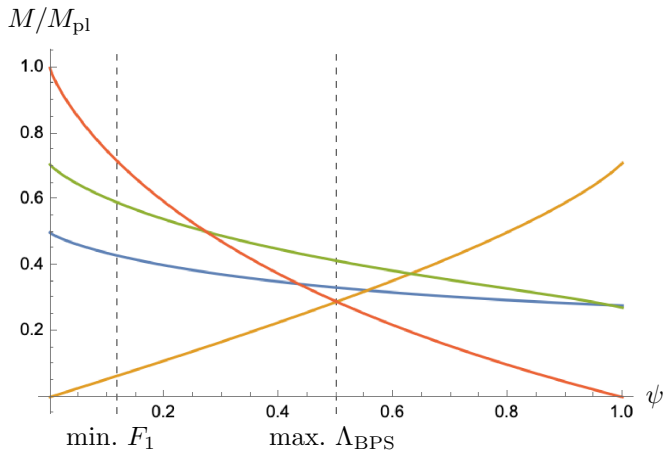

Having established a location for the desert, we next comment on the masses of BPS states at this point and in its vicinity. We focus on the states that become massless at the singularities in the moduli space: the D2-brane and D0-brane states at the large complex structure point, the D6-brane state at the conifold point at $x=1/1024$ and the conifold state for the point at $x=\infty$. The masses of these states at the minimum $F_1$ point, in increasing order, are
\begin{equation}\begin{aligned}
\frac{M_{\rm 2D6+D4+4D0}}{M_{\rm pl}} &= 0.0648039\, , \quad &\frac{M_{\rm D2}}{M_{\rm pl}} &= 0.430100\, ,\\
\frac{M_{\rm D0}}{M_{\rm pl}} &= 0.591146\, , \quad &\frac{M_{\rm D6}}{M_{\rm pl}} &= 0.716418
\end{aligned}\end{equation}
where the first, lightest mass corresponds to the state of the conifold point at $x=\infty$. Interestingly, for just the BPS states alone this would not constitute the desert point; in that case we could simply move further away from the conifold point at $x=\infty$ in order to maximize the gap. For illustration we have included a plot of the masses along the real line in between the conifold point at $x=\infty$ and $x=1/1024$ in figure~\ref{fig:6masses}.

\subsection{The mirror bicubic \texorpdfstring{$X_{3,3}(1^6)$}{X33}}\label{sec:X33}
In this section we consider again an example without an orbifold point in moduli space, in this case the mirror of the bicubic $X_{3,3}$ in $\mathbb{P}_{1,1,1,1,1,1}$. The purpose of this example is to consider a moduli space that has an emergent string limit, cf.~section \ref{sec:Kpoint}, which is realized at the K-point that is present in this moduli space. The moduli space contains three singularities: a large complex structure limit at $x=0$, a conifold point at $x=3^{-6}$ and a K-point at $x=\infty$. The monodromy around this K-point contains a semi-simple part of order three (the local periods have exponents $(\frac{1}{3},\frac{1}{3},\frac{2}{3},\frac{2}{3})$). The one-parameter variation of Hodge structure of this Calabi--Yau threefold corresponds to the hypergeometric family AESZ 4 in the database \cite{AESZ}. 

For this geometry the genus one free energy of the topological string is given by
\begin{equation}
    F_1 = 8 K -\frac{1}{2} \log[G_{x\bar x}] + \log |x^{-\frac{11}{2}}(1-3^6 x)^{-1/12}|^2\, .
\end{equation}
The holomorphic ambiguity has been fixed by requiring the correct asymptotics at the conifold point at $x=3^{-6}$ (see \eqref{asymptconi}) and the LCS point \eqref{asymptLCS} with $c_2=54$. For the K-point at $x=\infty$, in terms of the local coordinate given by $x=3^6 \psi^3$,\footnote{The cube assures that there is no finite order part for the monodromy in the local coordinate $\psi$.} $F_1$ diverges as
\begin{equation}
F_1 \stackrel{\hspace{-3pt}\text{\tiny K-point}}{\longrightarrow} -\log|\psi|^2\, .
\end{equation}
This asymptotic behaviour is consistent with our discussion of K-points and their analogues in higher-dimensional moduli spaces in section~\ref{sec:Kpoint}.

\begin{figure}[!ht]
\vspace*{2pt}
\centering

\begin{subfigure}{0.49\textwidth}
\centering
\vspace*{10pt}\includegraphics[width=230pt]{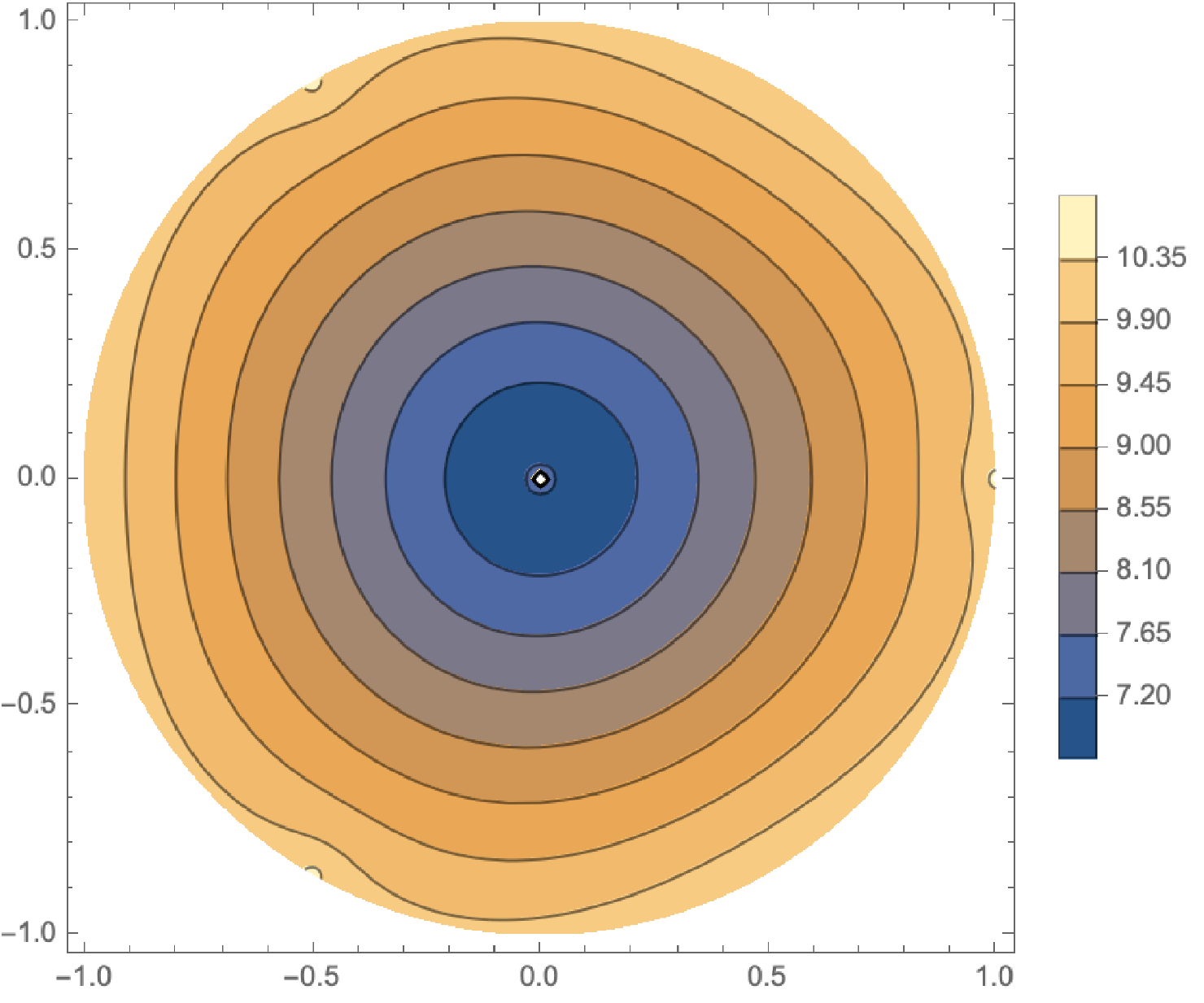}%
\vspace*{-20pt}\caption{K-point in $\psi= \frac{1}{9}x^{-1/3}$.\label{fig:4infconifold}}
\end{subfigure}
\hspace{2pt}
\begin{subfigure}{0.49\textwidth}
\centering
\vspace*{10pt}\includegraphics[width=230pt]{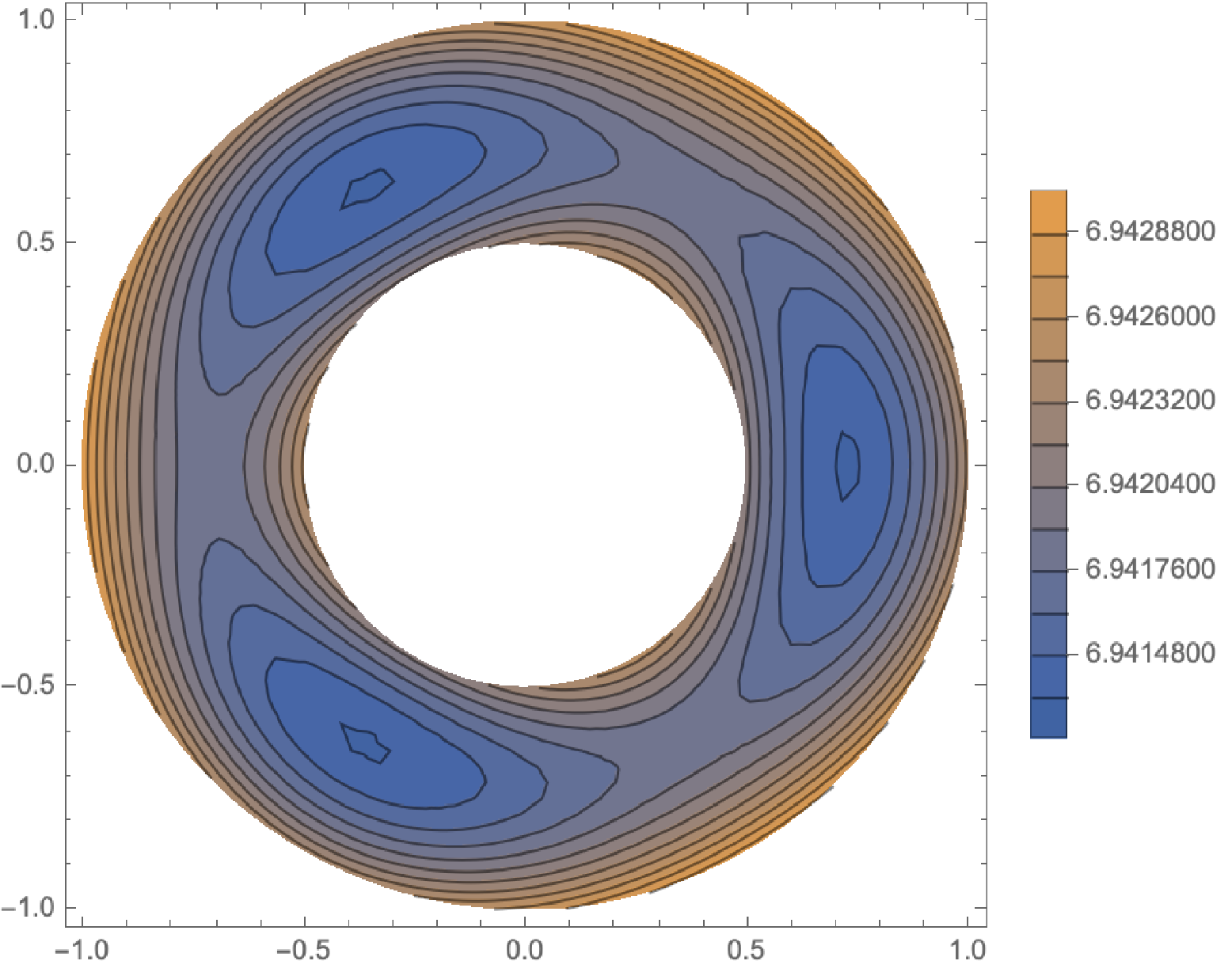}%
\vspace*{-20pt}\caption{K-point in $\psi_{\rm res}=\frac{ \psi}{200 |\psi|}(4|\psi|+15) $.\label{fig:4infconifoldzoom}}
\end{subfigure}
\caption{Contour plots of $F_1$ in the moduli space of $X_{3,3}$ in $\mathbb{P}_{1,1,1,1,1,1}$. For the K point at $x=\infty$ we included two figures: in \ref{fig:4infconifold} we see a band around the K-point where $F_1$ takes the lowest value; in \ref{fig:4infconifoldzoom} we rescale the coordinate to zoom in, from which it becomes clear that $F_1$ is minimized precisely on the straight lines between the K-point at $\psi=0$ and the conifold points at the third roots of unity. \label{fig:F1AESZ4}}
\end{figure}

Next we investigate where the species scale is maximized and thus $F_1$ is minimized. At all singularities $F_1$ diverges, so we again have to look elsewhere for this point. Similar to the previous example we find that the minimum occurs in the patch around the point at $x=\infty$, in this case the K-point. We find that $F_1$ takes the smallest values in a band around the K-point, as is illustrated by the contour plots in figure~\ref{fig:F1AESZ4}. Zooming in closer on this band, we see that there are three minima for $F_1$, related by the $\mathbb{Z}_3$-part of the monodromy around this singularity. Again they lie precisely on the lines between the conifold point at infinity and the other conifold points at the fourth roots of unity. For completeness, let us record the location in the Picard--Fuchs coordinate
\begin{equation}
x_{\rm desert} = 1.91538\, .
\end{equation}
There are no special algebraic relations among the periods compared to any points in its vicinity, so we do not include them here.

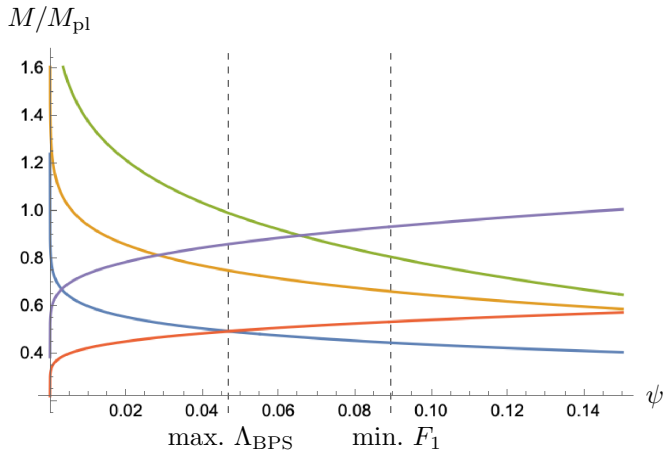
\begin{figure}[!ht]
\centering
\vspace*{10pt}
\includegraphics[width=230pt]{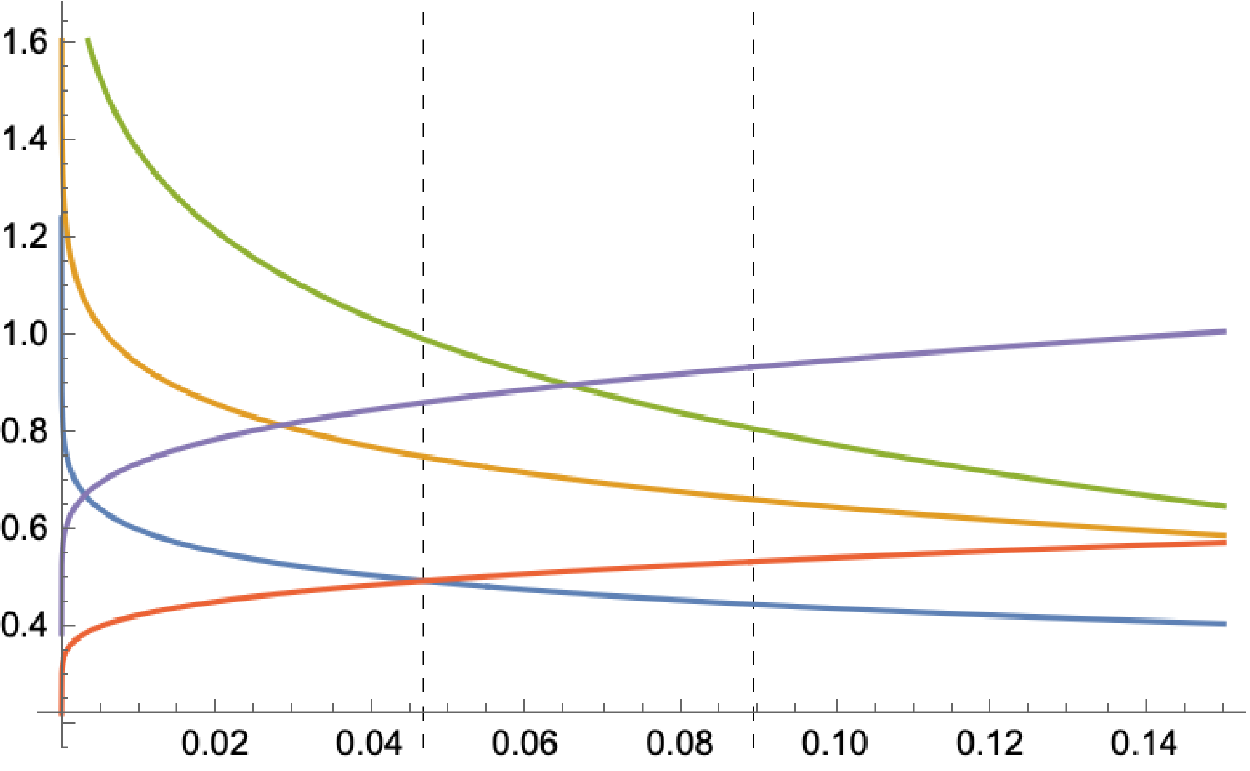}%
\begin{picture}(0,0)
\put(-235,150){\footnotesize$M/M_{\rm pl}$}
\put(4,8){\footnotesize$\psi$}
\put(-175,-8){\footnotesize max.~$\Lambda_{\rm BPS}$}
\put(-106,-8){\footnotesize min.~$F_1$}
\end{picture}\vspace{3mm}
\caption{Plot of the masses of the D2-brane (blue), D0-brane (yellow), D6-brane (green), D6+3D2-brane (red) and D4-5D2+3D0-brane (purple). We used the coordinate $\psi$ defined by $x=1/(9\psi)^3$, in which the K-point at $x=\infty$ is located at $\psi=0$ and the conifold at $x=1/3^6$ at $\psi=1$. The minimum of $F_1$ is located at $\psi=0.089469$, while the maximum gap of the BPS states in the above plot is to the left of this at $\psi=0.0468646$, both of which have been indicated by dashed lines.\label{fig:4masses}}
\end{figure}

Given this location for the desert, we now compare with the masses of the light BPS states. Focusing on the states that become massless at the singularities in moduli space, we consider: the dual of D2 and D0-brane states for the large complex structure point $x=0$, the dual of the D6-brane state at the conifold point $x=3^{-6}$ and the bound D-brane states dual to D6+3D2 and D4-5D2+3D0 associated to the K-point at $x=\infty$. Their masses at the minimum of $F_1$ are given by
\begin{equation}
\begin{aligned}
 \frac{M_{\rm D0}}{M_{\rm pl}}&=0.665849 \, , \qquad &\frac{M_{\rm D2}}{M_{\rm pl}}&= 0.449986\, ,  \quad &\frac{M_{\rm D6}}{M_{\rm pl}}&= 0.811345\, , \\
 \frac{M_{\rm D4-5D2+3D0}}{M_{\rm pl}}&=0.939319 \, , \qquad &\frac{M_{\rm D6+3D2}}{M_{\rm pl}} &= 0.538613\, .
\end{aligned}
\end{equation}
Thus the lightest BPS state at this desert point is the single D2-brane that becomes massless at the LCS point. Note, however, that there is again no other BPS state with the same mass, so we can achieve a larger BPS mass gap by moving away from this point. To illustrate this point we have included a plot of the masses along the real line between $x=\infty$ and $x=3^{-6}$ in figure~\ref{fig:4masses}. The largest BPS mass gap is located at $x=13.3271$. The periods at this point take a special form, so let us record them for completeness
\begin{equation}
    \Pi = (0.195427, \, 0, \,0.737388, \, 0)\, + 0.0860222 i\,  (2,\, 2, \, -1, \, 4)\, , 
\end{equation}
where we note that, after a suitable overall rescaling, only two real numerical constants remain. This structure is reminiscent of other special points in the moduli space such as rank-2 attractors \cite{Moore:1998pn,Candelas:2019llw, Candelas:2021mwz, Bonisch:2022slo} and vacua with a vanishing flux superpotential \cite{DeWolfe:2005gy,Kachru:2020sio, Kachru:2020abh}. It would be interesting to better understand this connection between algebraic relations among the periods and the maximal BPS gap.

\section{Conclusions}\label{sec:conclusions}
In this paper we have proposed an index-like expression for the moduli dependence of the number of light species $N_{\rm sp}$ for 4d ${\cal N}=2$ supergravity theories given by the one loop topological string free energy. We have motivated this both from the view point of the B-model computation of determinants of Laplacians on CY as well as the analog of an $a$-function in the supergravity context.  
We have checked that this proposal agrees with expectations at the boundaries of moduli space at large volume (LCS), conifold points, and emergent string points (K-points).  Moreover we have used this to find the desert points in moduli space where we have the least amount of light species and $\Lambda_{\rm sp}$ is maximized.
We have also compared this with the mass gap of BPS states and have shown that even though in some cases they agree, in general they do not agree.  Thus $F_1$ carries more information than just the BPS gap.

Even though our index-like proposal for the species scale passes some highly non-trivial checks, we have also pointed out a short-coming of it:  It is index-like, and so in some cases it does not properly account the light degrees of freedom because of boson-fermion cancellation. 

It would be interesting to try to define similar index-like species scales with different number of supersymmetries and in different dimensions.   Also it would be important to find a direct relation between the usual definition of species scale in terms of black hole physics with our index-like substitute for it.

\subsubsection*{Acknowledgments} 
We would like to thank Ignatios Antoniadis, Brice Bastian, Eduardo Garc\'ia-Valdecasas, Thomas Grimm, Miguel Montero, Lorenz Schlechter and John Stout for interesting discussions. The work of CV, MW, and DW is supported by a grant from the Simons Foundation (602883,CV) and by the NSF grant PHY-2013858.

\appendix

\section{Details on Calabi--Yau periods}\label{app:periods}
In this appendix we cover some of the details on the periods of the holomorphic $(3,0)$-form $\Omega$ of Calabi--Yau manifolds. We begin with a brief introduction where we review how to compute these periods using Picard--Fuchs methods. We then provide some explicit expressions for the examples we consider, where we record the monodromy matrices and write down the charges of the light BPS states at the singularities.

\subsection{Picard--Fuchs equations and transition matrices}
In this section we lay out the procedure to determine the periods in an integral symplectic basis in all possible patches of the moduli space. The starting point of our discussion is the Picard--Fuchs equation. In this work we will be concerned with variations of Hodge structure for hypergeometric families, in which case the differential operator reads 
\begin{equation}\label{eq:pf}
L=\theta^4 -  \frac{x}{x_c} (\theta+a_1)(\theta+a_2)(\theta+a_3)(\theta+a_4)\, ,
\end{equation}
where we defined $\theta = x \frac{\partial}{\partial x}$. The parameters $a_1,a_2,a_3,a_4$ and $x_c$ are determined by the choice of Calabi--Yau threefold. More specifically, the $a_i$ give the exponents of the local solutions around the singularity $x=\infty$. The other singularities are located at $x=x_c$ which is a conifold point, and $x=0$ which is a large complex structure point.

Here we set the conventions for the integral symplectic basis of our periods. This basis is determined by the asymptotic form of the periods in the large complex structure regime. In this limit the prepotential reads
\begin{equation}
\cF = -\frac{1}{6} \kappa t^3 - \frac{1}{2}\sigma t^2 +\frac{c_2}{24} t + \frac{ c_3 \zeta(3)}{2(2\pi i)^2} \, .
\end{equation}
Here $\kappa,\sigma,c_2,c_3$ are given by the topological data of the mirror manifold: $\kappa$ is the triple intersection number, $c_2,c_3$ the integrated second and third Chern classes, and $\sigma = (\kappa \mod 2)/2$. We straightforwardly compute the periods from this prepotential to be
\begin{equation}\label{eq:intbasis}
\Pi = \begin{pmatrix}
    1 \\
    t \\
    \partial_t \cF \\
    2\cF - t\partial_t \cF
\end{pmatrix} = \begin{pmatrix}
    1 \\
    t \\
    -\frac{\kappa}{2} t^2 - \sigma t +\frac{c_2}{24} \\
    \frac{\kappa}{6} t^3 +\frac{c_2}{24} t -\frac{ c_3 \zeta(3)}{(2\pi i)^3}
\end{pmatrix}\, .
\end{equation}
The instanton contributions to these periods can be determined by solving the Picard--Fuchs equation \eqref{eq:pf} by a power series ansatz for each of the periods. More specifically, we take a local basis of periods given by
\begin{equation}\label{eq:LCSseries}
\varpi_{\rm LCS} = \begin{pmatrix}
 f_0(x)  \\
f_0(x) \frac{\log[x]}{2\pi i} + f_1(x) \\
f_0(x) \frac{(\log[x])^2}{(2\pi i)^2} + 2 f_1(x) \frac{\log[x]}{2\pi i} + f_2(x)\\
f_0(x) \frac{\log[x]^3}{(2\pi i)^3}+3f_1(x)\frac{\log[x]^2}{(2\pi i)^2} + 3f_2(x)\frac{\log[x]}{2\pi i}   + f_3(x)
\end{pmatrix}\, ,
\end{equation}
where we define series ans\"atze for the analytic functions as
\begin{equation}
\begin{aligned}
    f_0(x) &= 1 + c_{01} x+ c_{02} x^2+\ldots\, , \\
    f_1(x) &= c_{11} x+ c_{12} x^2+\ldots\, , \\
    f_2(x) &= c_{21} x+ c_{22} x^2+\ldots\, , \\
    f_3(x) &= c_{31} x+ c_{32} x^2+\ldots\, , \\
\end{aligned}
\end{equation}
Note that we normalized these power series such that $f_0(0)=1$ and $f_1(0)=f_2(0)=f_3(0)=0$. By plugging this ansatz into the Picard--Fuchs equation \eqref{eq:pf} we can systematically determine the coefficients $c_{nm} \in \mathbb{Q}$. In general, however, these periods are in a complex basis rather than an integral basis. In order to rotate to the integral basis we can match the leading terms in $t=\log[x]/2\pi i$ with the periods in  \eqref{eq:intbasis}. This results in a basis transformation to the integral symplectic basis
\begin{equation}
\Pi = T_{\rm LCS} \varpi_{\rm LCS} \, .
\end{equation}
For the large complex structure point it can be solved in general to be
\begin{equation}\label{eq:TLCS}
    T_{\rm LCS} = \left(
\begin{array}{cccc}
 1 & 0 & 0 & 0 \\
 0 & 1 & 0 & 0 \\
 \frac{c_2}{24} & -\sigma  & -\frac{\kappa }{2} & 0 \\
 \frac{i c_3 \zeta (3)}{8 \pi ^3} & \frac{c_2}{24} & 0 & \frac{\kappa }{6} \\
\end{array}
\right)\, .
\end{equation}
Next we consider the periods near the other boundaries in moduli space, say some singularity at $x=x_s$. Using the same approach as before one can first define a local basis of periods $\varpi_s$ similar to \eqref{eq:LCSseries}. In order to rotate to the integral symplectic basis one next has to match with the basis of periods at the LCS point. If these have an overlapping region of convergence we can simply expand both sets of periods around a point in the middle $x = x_{m}$ up to fourth order, or evaluate at four different points in this overlap region. This is the case for the conifold point, where we match the periods by solving 
\begin{equation}
\frac{\partial^k}{\partial x^k} \Pi \big|_{x=x_c/2} =  T_c \cdot \frac{\partial^k}{\partial x^k}\varpi_c \big|_{x=x_c/2} \, ,
\end{equation}
for $k=0,1,2,3$ for the numerical\footnote{Recently there has been much progress \cite{Bonisch:2022mgw} in uncovering the arithmetic origin of these coefficients.} coefficients of the matrix $T_c$. On the other hand, for points at $x=\infty$ there is no overlap in the region of convergence with the LCS point. In this case the integral basis can for instance be inferred numerically from the basis for the conifold point periods by going to a mid-point $x=3x_c/2$ between these two singularities. In the end we always find a transition matrix from the local basis of periods $\varpi_s$ to the integral period vector $\Pi$ by some numerical transition matrix as
\begin{equation}
\Pi = T_{\rm s} \, \varpi_s\, .
\end{equation}
Given these techniques we can now compute the periods in the examples of interest to our work. In the following subsections we record the most important results about the periods for each of the examples, such as the monodromy matrices and light states associated to the singularities.

\subsection{Example: mirror quintic \texorpdfstring{$X_{5}(1^6)$}{X5}}
Here we write down the details on the period vector for the mirror quintic studied in section~\ref{sec:X5}. We first record the Picard--Fuchs operator that captures the complex structure moduli dependence of the periods. Its differential operator, see e.g.~\cite{Candelas:1990rm}, takes the form
\begin{equation}
L_{X_5} = \theta^4 - 5^5 x (\theta+\frac{1}{5})(\theta+\frac{2}{5})(\theta+\frac{3}{5})(\theta+\frac{4}{5})\,.
\end{equation}
It has a large complex structure singularity at $x=0$, a conifold point $x=5^{-5}$ and a Landau-Ginzburg point at $x=\infty$. In order to fix the integral basis for the periods we need the topological data of the quintic. The intersection number and second and third Chern classes are given by
\begin{equation}
\kappa = 5\, , \quad c_2 = 50\, , \quad c_3=\chi=-200\, .
\end{equation}
With this data at hand, we can analytically continue the periods towards the vicinity of every singularity in the moduli space. This allows us to read off the monodromies around these points as
\begin{equation}
M_0 = \scalebox{0.95}{$\left(
\begin{array}{cccc}
 1 & 0 & 0 & 0 \\
 1 & 1 & 0 & 0 \\
 -3 & -5 & 1 & 0 \\
 5 & 2 & -1 & 1 \\
\end{array}
\right)$}, \quad M_{5^{-5}} =\scalebox{0.95}{$\left(
\begin{array}{cccc}
 1 & 0 & 0 & -1 \\
 0 & 1 & 0 & 0 \\
 0 & 0 & 1 & 0 \\
 0 & 0 & 0 & 1 \\
\end{array}
\right)$}, \quad  M_\infty = \scalebox{0.95}{$\left(
\begin{array}{cccc}
 -4 & 3 & 1 & 1 \\
 -1 & 1 & 0 & 0 \\
 -2 & 5 & 1 & 0 \\
 -5 & 3 & 1 & 1 \\
\end{array}
\right)$}.
\end{equation}
In turn, we determine the light states near every singularity in the moduli space. For the large complex structure point we have the D2- and D0-brane states (and their bound states)
\begin{equation}
q_{\rm D2} = (0, \, 0 , \, 1 , \, 0)\, ,\qquad q_{\rm D0} = (0, \, 0 , \, 0 , \, 1)\, , 
\end{equation}
while for the conifold point we have the single D6-brane state
\begin{equation}
    q_{\rm D6} = (1, \, 0, \, 0, \, 0 )\, .
\end{equation}
At the Landau-Ginzburg point there are no associated light states, so studying these states suffices to determine the location of the BPS desert.

\subsection{Example: mirror of \texorpdfstring{$X_{4,2}(1^6)$}{X42}}
As our next example we study the mirror of the intersection, $X_{4,2}$, of a quartic and a quadric in $\mathbb{P}_{1,1,1,1,1,1}$. The Picard--Fuchs equation that captures the dependence of the periods on the complex structure moduli reads
\begin{equation}
L_{X_{4,2}} = \theta^4 - 2^{10} x(\theta+\frac{1}{4})(\theta+\frac{1}{2})^2(\theta+\frac{3}{4})\,.
\end{equation}
with the large complex structure point is located at $x=0$, the conifold point at $x=2^{-10}$ and another conifold point at $x=\infty$ (the local periods have exponents $(\frac{1}{4}, \frac{1}{2}, \frac{1}{2}, \frac{3}{4})$). In order to rotate to the integral basis via \eqref{eq:TLCS}, let us record the topological data of $X_{4,2}$, which is given by
\begin{equation}
\kappa = 8\, , \quad c_2 = 56\, , \quad c_3=\chi=-176\, .
\end{equation}
With this information we can compute the periods near each of the singularities in the moduli space. We find the monodromies around these points to be
\begin{equation}
M_0 = \scalebox{0.95}{$\left(
\begin{array}{cccc}
 1 & 0 & 0 & 0 \\
 1 & 1 & 0 & 0 \\
 -4 & -8 & 1 & 0 \\
 6 & 4 & -1 & 1 \\
\end{array}
\right)$}, \quad M_{2^{-10}} =\scalebox{0.95}{$\left(
\begin{array}{cccc}
 1 & 0 & 0 & -1 \\
 0 & 1 & 0 & 0 \\
 0 & 0 & 1 & 0 \\
 0 & 0 & 0 & 1 \\
\end{array}
\right)$}, \quad  M_\infty = \scalebox{0.95}{$\left(
\begin{array}{cccc}
 -5 & 4 & 1 & 1 \\
 -1 & 1 & 0 & 0 \\
 -4 & 8 & 1 & 0 \\
 -6 & 4 & 1 & 1 \\
\end{array}
\right)$}.
\end{equation}
The monodromy around the conifold point at infinity deserves some special attention. It may be decomposed into a semi-simple and a unipotent part as
\begin{equation}
M_\infty = M_\infty^u M_\infty^s\, , \quad M_\infty^u =\left(
\begin{array}{cccc}
 5 & 0 & -1 & -2 \\
 2 & 1 & -\frac{1}{2} & -1 \\
 0 & 0 & 1 & 0 \\
 8 & 0 & -2 & -3 \\
\end{array}
\right) \, , \quad M_\infty^s = \left(
\begin{array}{cccc}
 -1 & 4 & 0 & -1 \\
 1 & 1 & -\frac{1}{2} & -1 \\
 -4 & 8 & 1 & 0 \\
 2 & 4 & -1 & -3 \\
\end{array}
\right)\, .
\end{equation}
where the semi-simple part is of order four, i.e.~$(M_\infty^s)^4 = 1$. The state that becomes massless at this conifold point can be read off from the monodromy matrix as
\begin{equation}
q_\infty = (2,\,  1,\,  0,\, 4 ) \in \text{Img}(M^{u}_\infty-1)\, .
\end{equation}
We can write the fourth power of the monodromy as
\begin{equation}\label{eq:AESZ6modG}
    (M_\infty)^4 = 1 + 2 \, q_\infty (\Sigma q_\infty)^T\, ,
\end{equation}
which is the monodromy around the fourfold covering of this singularity without any semi-simple part. From the coefficient in this expression we can read off that the conifold cycle is quotiented by a finite group $G$ of order $|G|=2$. The other light states in the moduli space are, similar to the quintic, the D2- and D0-brane states at the LCS point and the D6-brane state at the first conifold point. 

\subsection{Example: mirror bicubic \texorpdfstring{$X_{3,3}(1^6)$}{X33}}
Finally we consider the mirror of the bicubic $X_{3,3}$ in $\mathbb{P}_{1,1,1,1,1,1}$. The complex structure moduli dependence of its periods is captured by the Picard--Fuchs operator
\begin{equation}
L_{X_{3,3}} = \theta^4 - 3^6 x (\theta+\frac{1}{3})(\theta+\frac{1}{3})(\theta+\frac{2}{3})(\theta+\frac{2}{3})\,.
\end{equation}
The corresponding complex structure moduli space has as singularities a large complex structure point at $x=0$, a conifold point at $x=3^{-6}$ and a K-point at $x=\infty$ with local exponents $(\frac{1}{3}, \frac{1}{3}, \frac{2}{3}, \frac{2}{3})$. The topological data that fixes the integral basis at the large complex structure point is given by
\begin{equation}
\kappa = 9\, , \quad c_2 = 54\, , \quad c_3=\chi=-144\, .
\end{equation}
With this information at our disposal, we can compute the periods numerically in an integral basis near every singularity in the moduli space. This allows us to read off the monodromy matrices associated to these points as
\begin{equation}
M_0 = \scalebox{0.95}{$\left(
\begin{array}{cccc}
 1 & 0 & 0 & 0 \\
 1 & 1 & 0 & 0 \\
 -5 & -9 & 1 & 0 \\
 6 & 4 & -1 & 1 \\
\end{array}
\right)$}\, , \quad M_{3^{-6}} = \scalebox{0.95}{$\left(
\begin{array}{cccc}
 1 & 0 & 0 & -1 \\
 0 & 1 & 0 & 0 \\
 0 & 0 & 1 & 0 \\
 0 & 0 & 0 & 1 \\
\end{array}
\right)$}\, , \quad M_\infty = \scalebox{0.95}{$\left(
\begin{array}{cccc}
 -5 & 5 & 1 & 1 \\
 -1 & 1 & 0 & 0 \\
 -4 & 9 & 1 & 0 \\
 -6 & 5 & 1 & 1 \\
\end{array}
\right)$}\, .
\end{equation}
The monodromy around the K-point at $x=\infty$ is deserving of some special attention. It may be decomposed into a semi-simple part of finite order and a unipotent part as
\begin{equation}
M_\infty = M^s_{\infty} M^u_{\infty}\, , \quad M^s_{\infty}=\left(
\begin{array}{cccc}
 -5 & 2 & 1 & 2 \\
 -2 & -\frac{1}{3} & \frac{1}{3} & 1 \\
 1 & \frac{20}{3} & -\frac{2}{3} & -2 \\
 -9 & 1 & 2 & 4 \\
\end{array}
\right)\, ,\quad M^u_{\infty} = \left(
\begin{array}{cccc}
 -2 & -1 & 1 & 2 \\
 -2 & \frac{4}{3} & \frac{2}{3} & 1 \\
 1 & -\frac{14}{3} & \frac{2}{3} & 1 \\
 -6 & 1 & 2 & 4 \\
\end{array}
\right),
\end{equation}
where $(M^s_{\infty})^3=0$. We can read off the states that become massless at the K-point from the unipotent part of the monodromy. We find bound states involving D4- and D6-branes given by
\begin{equation}
q_{\rm K1}, q_{\rm K2} \in \text{Img}(T_\infty^u -1): \qquad q_{\rm K1} =(0, \, 1, \, -5, \, 3)\, , \quad   q_{\rm K2} = (1, \,0, \, 3,\,  0)\, .
\end{equation}
In the mirror Type IIA setup the bound D4-brane state signals a 4-cycle becoming small in Planck units. It corresponds to the rigid K3 fiber in the geometry, which we expect can host a string arising from wrapping an NS5-brane. Sending the modulus to the K-point then results in an emergent string limit as discussed in more detail in section~\ref{sec:Kpoint}.

\bibliography{papers_Max}
\bibliographystyle{JHEP}

\end{document}